\documentclass[final,5p,times,twocolumn,authoryear]{elsarticle}

\usepackage{amsmath}
\usepackage{amssymb}
\usepackage{upgreek}

\usepackage{graphicx}
\usepackage{tikz}

\usepackage{caption}
\usepackage{subcaption}
\usepackage{makecell}

\usepackage{url}
\usepackage{hyperref}
\usepackage{breakurl}
\usepackage{cleveref}

\usepackage[english]{babel}
\usepackage[utf8]{inputenc}
\usepackage{dblfloatfix} 

\setcitestyle{square,comma,numbers}

\date{}

\begin{document}



\affiliation[inst1]{
  organization = {Department of Physics, Banaras Hindu University},
  city         = {Varanasi},
  postcode     = {221005},
  state        = {U.P.},
  country      = {India}
}

\author[inst1]{Rajiv Gupta}
\author[inst1]{Sunidhi Saxena}
\author[inst1]{Ajay Kumar\corref{cor1}}

\cortext[cor1]{Corresponding author. Email: ajay.phy@bhu.ac.in}

\title{Effect of hole pitch reduction on electron transport and diffusion: A comparative simulation study of Triple GEM detectors}

\begin{abstract}
\noindent
Advances in fabrication techniques and high-performance electronics have facilitated the development of fine-pitch Gas Electron Multipliers (GEMs). Earlier experimental and simulation findings suggest that these reduced-pitch GEMs can outperform the standard configuration in terms of effective gain, collection efficiency, and position resolution. However, a noticeable fraction of avalanche electrons is lost within the GEM systems, resulting in a degradation of charge collection efficiency. Therefore, a comprehensive simulation-based study is essential to provide deeper insights into the extent of degradation and its contributing factors. In this context, we employ ANSYS and Garfield++ to model the Triple GEM detectors with reduced pitch sizes of 90 and 60 $\mu$m, and perform a comparative performance analysis with the standard configuration (pitch size: 140 $\mu$m). At first, the simulation framework is validated by comparing the results of the standard configuration with available experimental data and previously reported simulation outcomes. Despite the characteristic gain offset, the framework remains physically consistent and reliable in capturing microscopic avalanche dynamics, reproducing the experimental trend. Following validation, we investigate electron losses at the metal electrodes and within the Kapton holes, electron transmission through the transfer and induction regions, electron diffusion on the induction electrode, and the overall collection efficiency. These parameters are analyzed as functions of GEM potential, outer hole diameter, inner hole diameter, Kapton thickness, metal thickness, and gas composition, thereby offering insights for designing efficient GEM detectors.

\end{abstract}
\maketitle

\section{Introduction}
%
%
%
%
Gaseous detectors \cite{Titov:2010br} operate on the principle of ionization initiated by the interaction of traversing charged particles (such as electrons, alpha particles, muons, etc.) or photons (X-rays) with the gaseous molecules. These particles create additional electron-ion pairs that drift in opposite directions due to the applied electric field. When the generated electrons come under the influence of a strong electric field, they ionize additional gas molecules, resulting in an avalanche effect. Consequently, enhanced signals are detected at the readout electrodes.  Over time, gaseous detectors have evolved from wire-based designs to advanced planar designs. Wire-based detectors, such as the Proportional Counter \cite{Rose:1941tdv, Charpak:1968kd}, Multi-Wire Proportional Chamber (MWPC) \cite{Walenta:1971hp}, and Drift Chambers \cite{Charpak:1973mug, Majewski:1983kna, Oh:1991kx}, use arrays of closely spaced (typically $\sim$mm) anode wires enclosed within the gas-filled chamber. The electric field is highly intense in the vicinity of the wires, where ionizing electrons trigger a Townsend avalanche. These detectors exhibit certain limitations, including low rate capability, accumulation of positive ions, limited multi-track resolution, mechanical fragility, aging effects, and non-uniform amplification. Many of these limitations can be overcome by employing planar structures, which achieve amplification through micro-patterned elements, such as holes, wells, or thin amplification gaps beneath a micromesh. These modern planar gas detectors are collectively referred to as Micro-Pattern Gaseous Detectors (MPGDs) \cite{Sauli:1999ya, Shekhtman:2002ih, Titov:2013hmq} and include technologies such as the Micro-Strip Gas Chamber (MSGC) \cite{Oed:1988jh, Oed:1995ku}, Micro-Mesh Gaseous Structure (MICROMEGAS) \cite{Giomataris:1995fq, Charpak:2001tp, Giomataris:2004aa}, Micro Compteur à Trous (MICROCAT) \cite{SARVESTANI1998238, Sarvestani:1999up},  micro-Resistive WELL ($\mu$-RWELL) \cite{Bencivenni:2014exa, PoliLener:2016pgy, Bencivenni:2017wee}, and Gas Electron Multiplier (GEM) \cite{ Sauli:1997qp, Bachmann:1999xc,  Sauli:2016eeu}.

\begin{figure*}[!ht]
\begin{center}
\begin{tikzpicture}

\filldraw[fill=blue!15, draw=black, thick] (-2,-5.25) rectangle (6,2.5);

\filldraw[fill=red!80, draw=black, thick] (-2,-5.25) rectangle (6,-5.15);

\filldraw[fill=pink!40, draw=black, thick] (1,-3.25) -- (3,-3.25) -- (3.25,-3.0) -- (3,-2.75) -- (1,-2.75) -- (0.75,-3.0) -- cycle;
\filldraw[fill=red!80, draw=black, thick] (1,-3.25) rectangle (3,-3.35);
\filldraw[fill=red!80, draw=black, thick] (1,-2.75) rectangle (3,-2.65);

\filldraw[fill=pink!40, draw=black, thick] (1,-1.5) -- (3,-1.5) -- (3.25,-1.25) -- (3,-1.0) -- (1,-1.0) -- (0.75,-1.25) -- cycle;
\filldraw[fill=red!80, draw=black, thick] (1,-1.5) rectangle (3,-1.60);
\filldraw[fill=red!80, draw=black, thick] (1,-1.0) rectangle (3,-0.90);

\filldraw[fill=pink!40, draw=black, thick] (1,0.25) -- (3,0.25) -- (3.25,0.5) -- (3,0.75) -- (1,0.75) -- (0.75,0.5) -- cycle;
\filldraw[fill=red!80, draw=black, thick] (1,0.25) rectangle (3,0.15);
\filldraw[fill=red!80, draw=black, thick] (1,0.75) rectangle (3,0.85);

\filldraw[fill=pink!40, draw=black, thick] (-2,-3.25) -- (-1,-3.25) -- (-0.75,-3.0) -- (-1,-2.75) -- (-2,-2.75) -- cycle;
\filldraw[fill=red!80, draw=black, thick] (-2,-3.25) rectangle (-1,-3.35);
\filldraw[fill=red!80, draw=black, thick] (-2,-2.75) rectangle (-1,-2.65);

\filldraw[fill=pink!40, draw=black, thick] (-2,-1.5) -- (-1,-1.5) -- (-0.75,-1.25) -- (-1,-1.0) -- (-2,-1.0) -- cycle;
\filldraw[fill=red!80, draw=black, thick] (-2,-1.5) rectangle (-1,-1.60);
\filldraw[fill=red!80, draw=black, thick] (-2,-1.0) rectangle (-1,-0.90);

\filldraw[fill=pink!40, draw=black, thick] (-2,0.25) -- (-1,0.25) -- (-0.75,0.5) -- (-1,0.75) -- (-2,0.75) -- cycle;
\filldraw[fill=red!80, draw=black, thick] (-2,0.25) rectangle (-1,0.15);
\filldraw[fill=red!80, draw=black, thick] (-2,0.75) rectangle (-1,0.85);

\filldraw[fill=pink!40, draw=black, thick] (5,-3.25) -- (6,-3.25) -- (6,-2.75) -- (5,-2.75) -- (4.75,-3.0) -- cycle;
\filldraw[fill=red!80, draw=black, thick] (5,-3.25) rectangle (6,-3.35);
\filldraw[fill=red!80, draw=black, thick] (5,-2.75) rectangle (6,-2.65);

\filldraw[fill=pink!40, draw=black, thick] (5,-1.5) -- (6,-1.5) -- (6,-1.0) -- (5,-1.0) -- (4.75,-1.25) -- cycle;
\filldraw[fill=red!80, draw=black, thick] (5,-1.5) rectangle (6,-1.6);
\filldraw[fill=red!80, draw=black, thick] (5,-1.0) rectangle (6,-0.90);

\filldraw[fill=pink!40, draw=black, thick] (5,0.25) -- (6,0.25) -- (6,0.75) -- (5,0.75) -- (4.75,0.5) -- cycle;
\filldraw[fill=red!80, draw=black, thick] (5,0.25) rectangle (6,0.15);
\filldraw[fill=red!80, draw=black, thick] (5,0.75) rectangle (6,0.85);

\filldraw[fill=red!80, draw=black, thick] (-2,2.5) rectangle (6,2.60);

\draw[thick] (9,-5.25) -- (9,2.60);
\draw[thick] (8.9,2.60) -- (9.1,2.60);
\draw[thick] (8.9,2.50) -- (9.1,2.50);
\draw[thick] (8.9,0.75) -- (9.1,0.75);
\draw[thick] (8.9,0.85) -- (9.1,0.85);
\draw[thick] (8.9,0.15) -- (9.1,0.15);
\draw[thick] (8.9,0.25) -- (9.1,0.25);
\draw[thick] (8.9,-0.9) -- (9.1,-0.9);
\draw[thick] (8.9,-1.0) -- (9.1,-1.0);
\draw[thick] (8.9,-1.5) -- (9.1,-1.5);
\draw[thick] (8.9,-1.6) -- (9.1,-1.6);
\draw[thick] (8.9,-2.65) -- (9.1,-2.65);
\draw[thick] (8.9,-2.75) -- (9.1,-2.75);
\draw[thick] (8.9,-3.25) -- (9.1,-3.25);
\draw[thick] (8.9,-3.35) -- (9.1,-3.35);
\draw[thick] (8.9,-5.15) -- (9.1,-5.15);
\draw[thick] (8.9,-5.25) -- (9.1,-5.25);

\node[right] at (6.2,3.5) {\small Config.1};
\node[right] at (6.2,3) {\small Config.2};
\node[right] at (7.55,3) {\small Config.3};
\draw[dashed] (7.52,2.7) -- (7.5,3.7);
\draw[dashed] (7.45,-5.8) -- (7.45,-5.2);
\draw[dashed] (10,2.7) -- (10,3.7);
\draw[dashed] (10.1,-5.8) -- (10.1,-5.3);

\node[right] at (9.1,2.60) {\small 4.120/4.120 mm};
\node[left] at (9.0,2.50) {\small 4.115/4.115 mm};
\node[right] at (9.1,0.75) {\small 2.110/2.085 mm};
\node[left] at (9.0,0.85) {\small 2.115/2.090 mm};
\node[right] at (9.1,0.15) {\small 2.055/2.055 mm};
\node[left] at (9.0,0.25) {\small 2.060/2.060 mm};
\node[right] at (9.1,-0.9) {\small 0.055/0.030 mm};
\node[left] at (9.0,-1.0) {\small 0.050/0.025 mm};
\node[right] at (9.1,-1.5) {\small 0/0 mm};
\node[left] at (9.0,-1.6) {\small -0.005/-0.005 mm};
\node[right] at (9.1,-2.65) {\small -2.005/-2.030 mm};
\node[left] at (9.0,-2.75) {\small -2.010/-2.035 mm};
\node[right] at (9.1,-3.25) {\small -2.060/-2.060 mm};
\node[left] at (9.0,-3.35) {\small -2.065/-2.065 mm};
\node[right] at (9.1,-5.25) {\small -5.070/-5.070 mm};
\node[left] at (9.0,-5.15) {\small -5.065/-5.065 mm};

\draw[thick] (-4,-5.25) -- (-4,2.60);
\draw[thick] (-4.1,2.60) -- (-3.9,2.60)  node[left,xshift=-0.2cm] {$V_{8}=0V$};

\draw[thick] (-4.1,0.85) -- (-3.9,0.85) node[left,xshift=-0.2cm] {$V_{7}= -700V$};;
\draw[thick] (-4.1,0.25) -- (-3.9,0.25) node[left,xshift=-0.2cm] {$V_{6}= -1050V$};;
\draw[thick] (-4.1,-1.0) -- (-3.9,-1.0) node[left,xshift=-0.2cm] {$V_{5}= -1750V$};
\draw[thick] (-4.1,-1.6) -- (-3.9,-1.6) node[left,xshift=-0.2cm] {$V_{4}$= -2100V};;
\draw[thick] (-4.1,-2.75) -- (-3.9,-2.75) node[left,xshift=-0.2cm] {$V_{3}= -2800V$};
\draw[thick] (-4.1,-3.35) -- (-3.9,-3.35) node[left,xshift=-0.2cm] {$V_{2}$= -3150V};
\draw[thick] (-4.1,-5.25) -- (-3.9,-5.25) node[left,xshift=-0.2cm] {$V_{1}= -3750$};

\node at (2,2.0) {\textbf{Ar:$\textbf{CO}_{2}$(70:30)}};
\draw[<->, thick] (-1,-1.6) -- (1,-1.6);
\node at (0,-1.8) {\textbf{\textsf{\scriptsize{70/55/30 \textmu m}}}};
\node at (0,-1.0) {\textbf{\textsf{\scriptsize{50/40/25 \textmu m}}}};
\draw[<->, thick] (-0.75,-1.25) -- (0.75,-1.25);
\node at (2,-3.5) {\textbf{\scriptsize{LM1}}};
\node at (2,-2.5) {\textbf{\scriptsize{UM1}}};
\node at (2,-1.75) {\textbf{\scriptsize{LM2}}};
\node at (2,-0.75) {\textbf{\scriptsize{UM2}}};
\node at (2,0.) {\textbf{\scriptsize{LM3}}};
\node at (2,1.0) {\textbf{\scriptsize{UM3}}};
\draw[<->, thick] (0,-0.25) -- (4,-0.25);
\node at (2,-0.42) {\textbf{\textsf{\scriptsize{140/90/60 \textmu m}}}};
\node at (0,-4.0) {\textbf{\scriptsize{Drift Region (3mm; 2kV/cm)}}};
\node at (0,-2.25) {\textbf{\scriptsize{Tran. 1 (2mm; 3.5kV/cm)}}};
\node at (0,-0.65) {\textbf{\scriptsize{Tran. 2 (2mm; 3.5kV/cm)}}};
\node at (1.0,1.5) {\textbf{\scriptsize{Tran. 3 [Induction Region] (2mm; 3.5kV/cm)}}};
\node at (2,-1.25) {\textbf{\textsf{\scriptsize{50/50/25 \textmu m}}}};
\node at (2,-5.0) {\textbf{\scriptsize{Drift electrode}}};
\node at (2,2.8) {\textbf{\scriptsize{Induction electrode}}};
\node at (5.5,-3) {\textbf{\scriptsize{GEM1}}};
\node at (5.5,-1.25) {\textbf{\scriptsize{GEM2}}};
\node at (5.5,0.5) {\textbf{\scriptsize{GEM3}}};

\end{tikzpicture}
\caption{Schematic of a modeled triple GEM detector showing three vertically stacked GEM foils with drift, transfer, and induction regions. Config. 1 uses 3 SGEMs, Config. 2 uses 3 FGEMs, and Config. 3 uses 3 FTGEMs. The left line indicates applied potentials, while the right line shows z-axis coordinates. The left side of the dotted line represents common z-axis coordinates of Configs. 1 and 2, while the right side denotes the z-axis coordinates of Config. 3. Electrons are accelerated upward from the bottom drift region to the Tran. 3 (induction) region. LM1, LM2, and LM3 denote the lower metal layers of GEM1, GEM2, and GEM3, respectively, while UM1, UM2, and UM3 represent the corresponding upper metal layers. Tran. 1, Tran. 2, and Tran. 3 indicate the transfer regions. The illustrated figure shows the operation of GEM detectors at $\Delta V_{\mathrm{GEM}}$= 350 V. 
}
\label{TripleGEM}
\end{center}
\end{figure*}
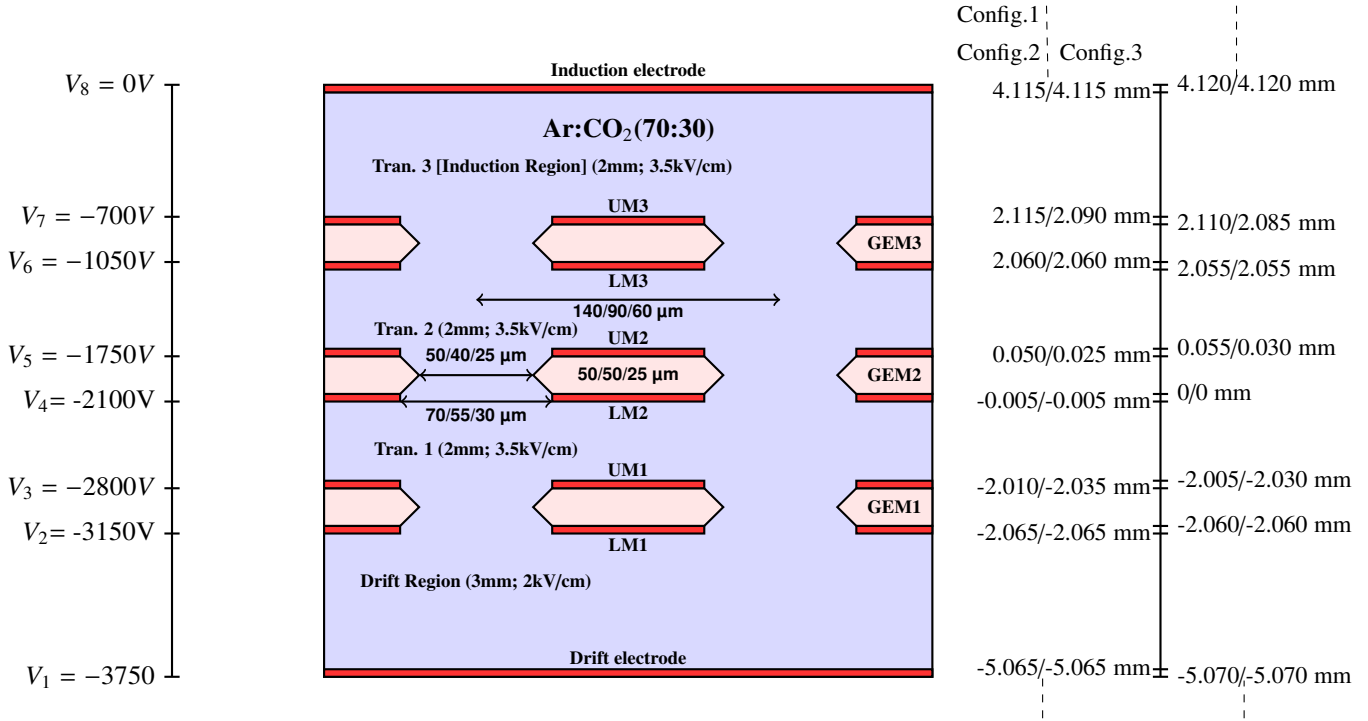

\begin{table*}[ht]
\centering
\scriptsize
\begin{tabular}{c|c|c|c}
\noalign{\vskip 4pt}
\hline

\textbf{Geometrical Parameters} & \makecell{\textbf{Config. 1} \\ 3SGEMs} & \makecell{\textbf{Config. 2} \\ 3FGEMs} & \makecell{\textbf{Config. 3} \\ 3FTGEMs} \\ 
\hline
Hole Pitch [$\mu$m] & 140 & 90 & 60 \\ 
Kapton Thickness [$\mu$m] & 50 & 50 & 25 \\ 
Metal layer [$\mu$m] & 5 & 5 & 5 \\ 
Inner diameter [$\mu$m] & 50 & 40 & 25 \\ 
Outer diameter [$\mu$m] & 70 & 55 & 30 \\ 
Drift gap [mm] & 3 & 3 & 3 \\
Drift electric field [kV/cm] & 2 & 2 & 2 \\
Transfer 1 / Transfer 2 / Transfer 3 (Induction) gap [mm] & 2 & 2 & 2 \\
Transfer 1 / Transfer 2 / Transfer 3 electric field [kV/cm] & 3.5 & 3.5 & 3.5 \\
 
\hline

\end{tabular}
\caption{Geometrical and electric field parameters used in the modeling of triple GEM detector configurations.}\label{table1}
\end{table*}
In 1997, F. Sauli introduced the GEM detector \cite{Sauli:1997qp}, a novel MPGD design constructed from a copper-clad Kapton foil and a dense array of biconical holes. These foils are positioned between the drift and induction regions and typically use a gas mixture of Argon and CO$_{2}$ in a 70:30 ratio. Argon acts as a primary ionizing gas, while CO$_{2}$ works as a photon quencher. When a suitable voltage is applied across the top and bottom GEM electrodes, a strong electric field is generated within the holes, resulting in charge amplification. Since the GEM serves exclusively as an amplification stage and is physically separated from the readout electrode, discharges are less likely to impact the front-end electronics. The overall gain of a GEM detector can be significantly enhanced by cascading multiple GEM foils along the electron drift direction. These cascaded foils reduce the discharge probability due to their effectiveness in low-voltage operation. They also exhibit desirable features, such as operational stability, radiation hardness, wide-area coverage, high counting rates, and good position resolution \cite{Ketzer:2001dt}. Consequently, they are adopted in most of the high-energy and nuclear physics experiments such as the Large Hadron Collider (LHC) \cite{Buonsante:2025lvd}, the Relativistic Heavy Ion Collider (RHIC) \cite{Simon:2008qk}, the Common Muon and Proton Apparatus for Structure and Spectroscopy (COMPASS) \cite{Ketzer:2004jk}, the Compressed Baryonic Matter (CBM) experiment \cite{Dubey:2013hva}, and the TOTal Elastic and diffractive cross-section Measurement (TOTEM) \cite{Lami:2006bd}. 

Future high luminosity physics experiments, such as the High Luminosity Large Hadron Collider (HL-LHC)~\cite{Hoepfner:2024vse} and the Electron-Ion Collider (EIC)~\cite{AbdulKhalek:2021gbh} will offer enhanced peak instantaneous luminosity of the order $10^{34}~\text{cm}^{-2}\text{s}^{-1}$. This enhanced luminosity will be beneficial in exploring rare physics events. To operate effectively under such intense radiation and ensure long-term stability, detector systems should be efficient, necessitating the optimization of GEM-based technologies. Recent experimental studies \cite{Flothner:2024qev, EPDT_Seminar, cern2023seminar} indicate that the reduced hole pitch GEM detectors can achieve enhanced gain and position resolution compared to standard configuration (as shown in Table \ref{table1}, Configuration 1 \cite{Bachmann:1999xc}). Our previous simulation study on a single GEM \cite{Gupta:2025yrv} also supported that decreasing pitch size and hole diameter can enhance the effective gain, collection efficiency, and sampling of the electron cloud. Additionally, it also indicated that reducing pitch size results in decreased electron transparency. Therefore, investigating the impact of pitch size on gain, collection efficiency, and transparency, along with the spatial spread of electrons on the induction plane, is essential for optimizing multi-layered GEM configurations.

Further, experimental setups are inherently limited in evaluating electron losses within different regions of a GEM system, as they typically rely on the final signal collection at the readout electrode without access to localized charge loss information. The charge collection at the readout could be significantly influenced by electron trapping within the GEM holes, absorptions at the GEM metals, and electron attachment to electronegative molecules \cite{Sauli:2016eeu, Alfonsi:2012msp, Tikhonov:2002um, Bouhali:2021vhg}. These charge losses can be effectively investigated and addressed through simulation-based studies. In this present study, we investigate these electron losses and their transmission in the transfer regions for three distinct experimentally investigated triple GEM configurations \cite{Flothner:2024qev, EPDT_Seminar, cern2023seminar}. Their specific dimensions and operating electric fields are listed in Table \ref{table1}. Since reducing the hole pitch enhances electron sampling, each configuration is investigated for the spread of electrons on the induction plane under various operating conditions. Moreover, the number of electrons collected at the readout plane is significantly impacted by the collection efficiency of the first GEM. It is defined as the probability that an electron transfers into the first GEM hole and initiates an avalanche without being lost to the lower metal surfaces. Therefore, each configuration is also investigated for its collection efficiency. Furthermore, various geometrical parameters, such as outer hole diameter, inner hole diameter, metal thickness, and Kapton thickness, as well as Ar-CO$_2$ gas compositions, are explored to understand their impact associated with pitch size reduction.

This article is organized as follows: Section \ref{section2} outlines the simulated detector geometries and the simulation framework. Section \ref{methods} describes the methodology utilized in the study. Section \ref{correlation} provides validation of the simulation framework. Section \ref{vgem} to \ref{gasconcentration} examines how changes in GEM potential, geometric design, and gas composition impact electron transport and their spread. Finally, section \ref{summary} summarizes and concludes the work.

\section{Geometric Modeling and Simulation Framework \label{section2}}

The simulation analysis is performed for three distinct GEM configurations (Config. 1, Config. 2, and Config. 3), as summarized in Table \ref{table1}. Each configuration has a triple-layered GEM with similar GEM foils arranged along the z-axis. Config. 1 utilizes three Standard Pitch GEMs (SGEMs), Config. 2 consists of three Fine Pitch GEMs (FGEMs), and Config. 3 employs three Fine Thin Pitch GEMs (FTGEMs). These configurations are motivated by experimental studies \cite{Flothner:2024qev, EPDT_Seminar, cern2023seminar}, which show improvement in position resolution with decreasing pitch size. The atmospheric pressure and temperature in this study are set to 1 atm and 293K, respectively. 

Fig. \ref{TripleGEM} shows the schematic cross-sectional view of the modeled GEM configurations, providing information about the dimensions of each region and the associated field distribution within the GEM system. On the left side of the figure, the applied voltages at each electrode are shown, demonstrating how desirable electric fields are maintained in the drift and transfer regions. The right side of the figure illustrates the node positions along the z-axis. It is worth noting that Config. 1 and Config. 2 have the same z-axis coordinates. In contrast, Config. 3 has different coordinates because of changes in Kapton thickness. All configurations are modeled using ANSYS \cite{ANSYS}, a finite element analysis software that precisely simulates the electrostatic behavior of the detector structures.

We employ ANSYS to model three unit cells embedded with metal contacts on either side, along with drift and induction electrodes, ensuring precise dimensions and positioning. The material properties for each component are defined appropriately. These unit cells are enclosed in a gas chamber filled with an Ar–CO$_{2}$ (70:30) mixture. The GEM configuration is discretized into meshes using SOLID123 tetrahedral elements. The hole regions and GEM foils are finely meshed due to the underlying field gradients, whereas the gas volume is meshed coarser to reduce computational load. The smart meshing feature helps automatically adjust element size based on geometric curvature and field gradient. All the volumes are glued together to ensure smooth and continuous meshing. The electric potential and electric field distribution are obtained at each mesh node by solving Maxwell's equations in the electrostatic limit. In addition, the numerical solutions provide element connectivity, nodal coordinates, and material properties. These output results are subsequently imported into Garfield++ \cite{Veenhof:1998tt} for further simulation and analysis. 

Garfield++ is an advanced simulation toolkit used for modeling gaseous and semiconductor detectors. It is widely used in the development of MPGDs and features updated electron transport properties. The unit cell designed in ANSYS can be mirrored in Garfield++ to simulate large detectors, due to its axisymmetric design. We have employed Magboltz \cite{Biagi:1989rm, Biagi:1999nwa} to solve the electron transport equations within the detector medium. Meanwhile, the microscopic tracking of electrons is simulated using a "AvalancheMicroscopic" module. However, it is important to note that the ANSYS-Garfield++ framework exhibits certain limitations. The simulation framework generally underestimates the experimental gain, primarily due to the absence of a photon feedback mechanism \cite{Sahin:2014haa}. Other contributing factors may include Magboltz's assumption of a uniform electric field \cite{Biagi:1999nwa}, the exclusion of space charge effects \cite{Rout:2021mbn}, surface charging of Kapton foil, limitations offered by the finite element solver, and low resolution for short-time microscopic interactions \cite{Jung:2021cvz}. Despite these intrinsic limitations, the combination of ANSYS and Garfield++ remains the most effective tool for simulating the microscopic behavior of GEM detectors. The presented results represent the ideal electrostatic response of the modeled GEM detectors, since only a static electric field and microscopic tracking have been considered.

\section{Simulation Methodologies \label{methods}}

The Monte Carlo simulation is initiated with a single electron of 0.5 eV initial energy, placed 0.5 mm below the drift electrode. Moreover, the electron's starting positions in the XY plane are randomized based on the pitch of the detector. The tracked electrons approach constant statistics at approximately 1000 events; therefore, the simulation results reflect the mean over such events. All three stacked GEM foils are supplied with the same potential difference, denoted as $\Delta V_{\mathrm{GEM}}$. The avalanche electrons drift through Tran. 1, Tran. 2, and Tran. 3 (induction) region before being collected at the induction electrode. A reasonable fraction of avalanche electrons reach the subsequent GEM holes, while others are lost due to metal absorptions, trapping within the GEM holes, and attachment in the transfer regions. Due to statistical fluctuation arising from Monte Carlo sampling, the collected electron count may exhibit a variation of approximately 2-3\%.

\begin{figure*}[t!]
    \centering
    \begin{subfigure}[b]{0.3\textwidth} 
        \centering
        \includegraphics[width=\textwidth, height=5cm]{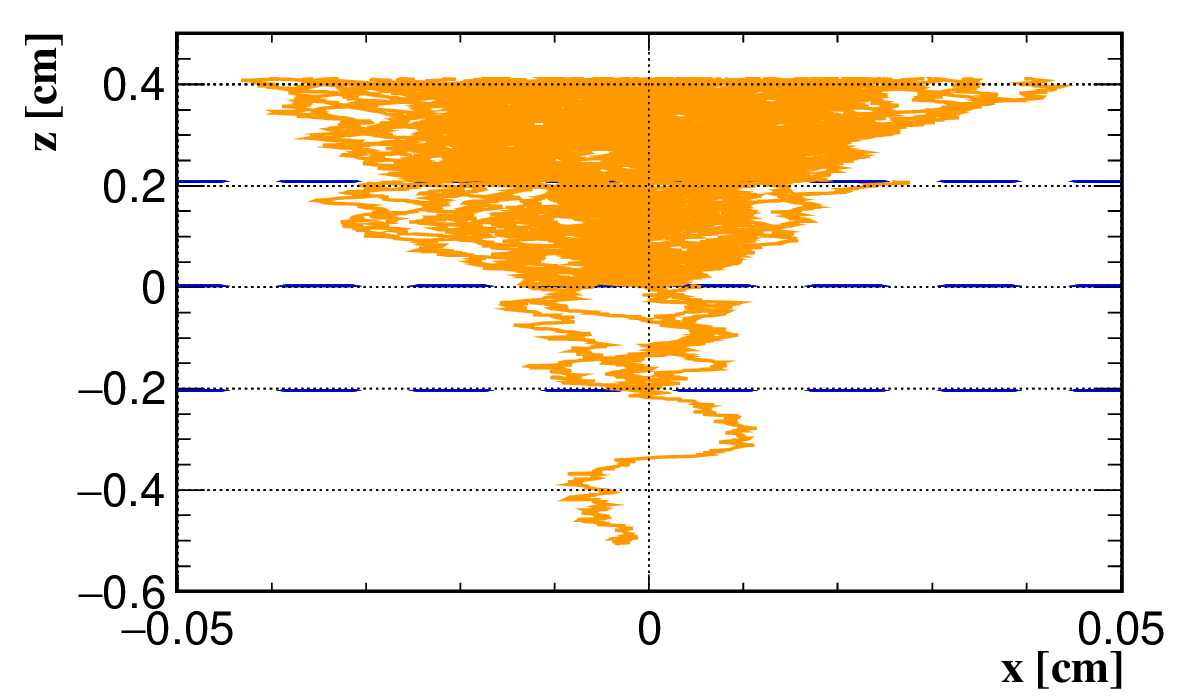}  
        \caption{}
        \label{fig:dlsgem}
    \end{subfigure}
    \hfill
    \begin{subfigure}[b]{0.3\textwidth}  
        \centering
        \includegraphics[width=\textwidth, height=5cm]{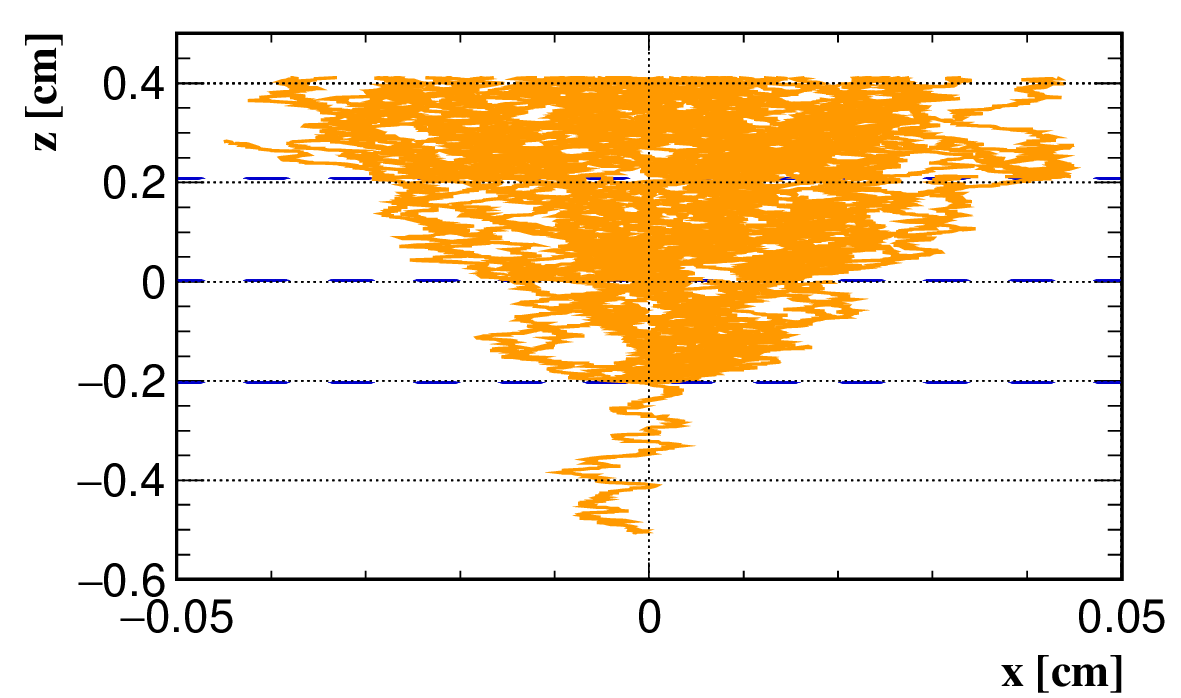}
        \caption{}
         \label{fig:dlfgem}
    \end{subfigure}
    \hfill
    \begin{subfigure}[b]{0.3\textwidth}  
        \centering
        \includegraphics[width=\textwidth, height=5cm]{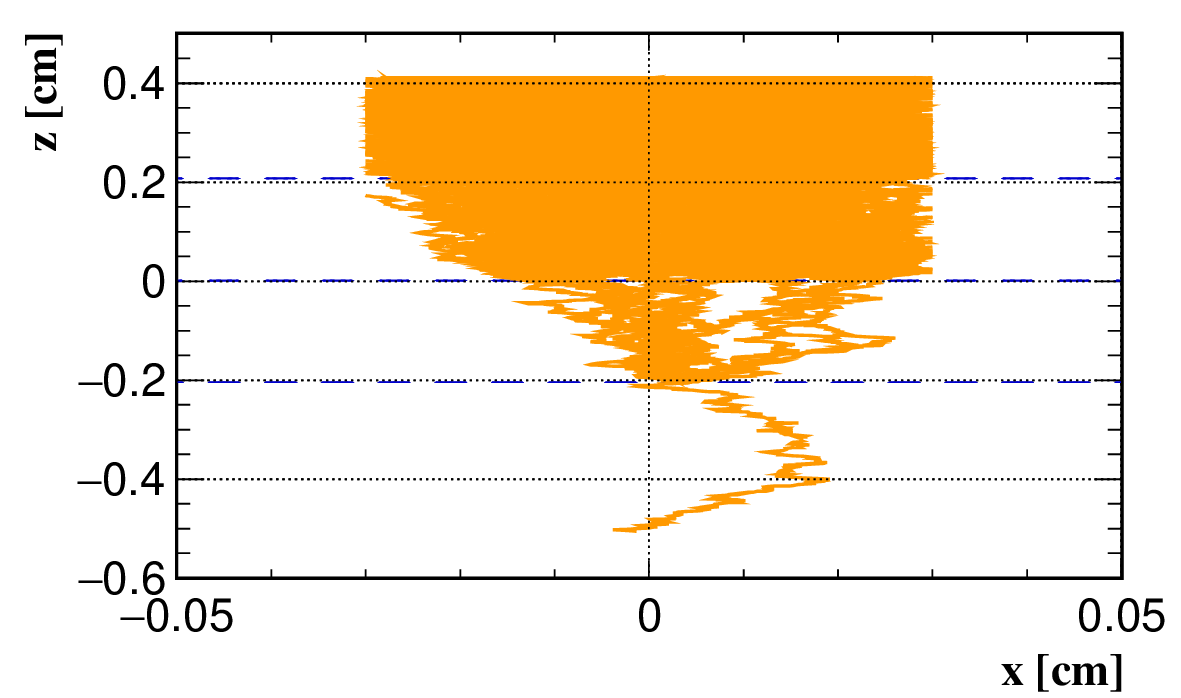}
        \caption{}
        \label{fig:dlftgem}
    \end{subfigure}

    \caption{Simulated drift lines initiated by a single electron at $\Delta V_{\mathrm{GEM}}$ of 300 V, displaying 2D (XZ) distributions of avalanche electrons in (a) Config. 1 (3 SGEMs), (b) Config. 2 (3 FGEMs), and (c) Config. 3 (3 FTGEMs).}
    \label{dlgem}
    
\end{figure*}

\begin{figure*}[t!]
    \centering
    \begin{subfigure}[b]{0.3\textwidth} 
        \centering
        \includegraphics[width=\textwidth, height=5cm]{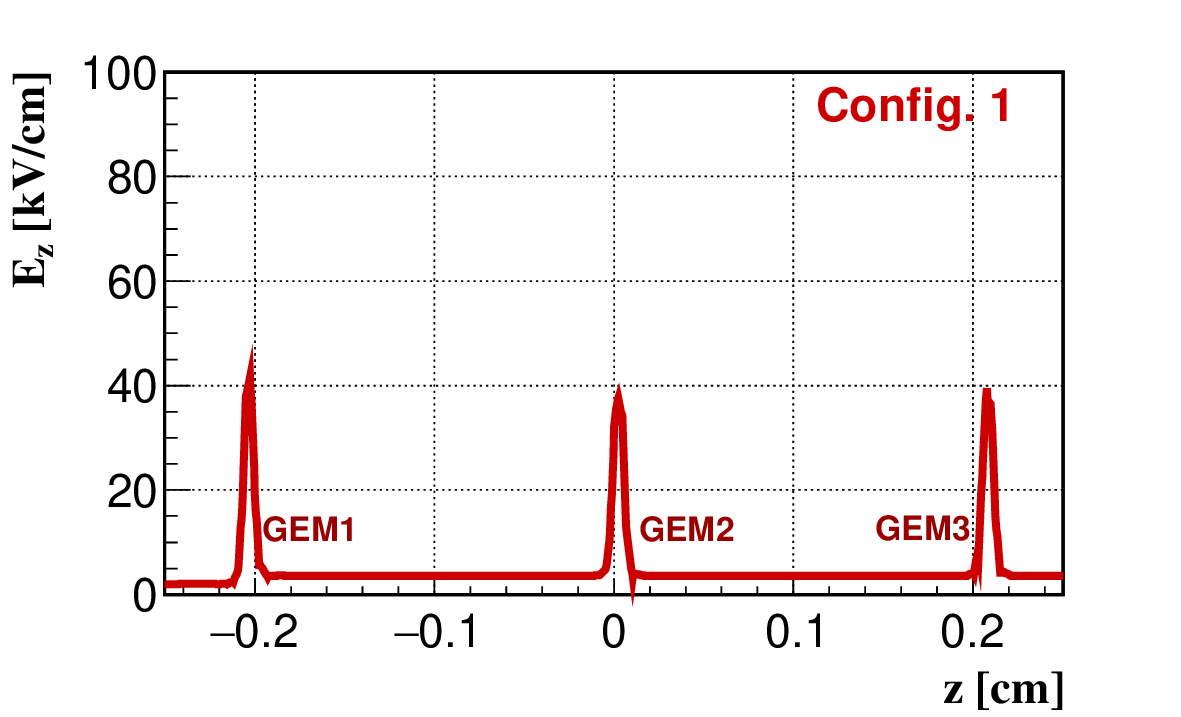}  
        \caption{}
        \label{fig:fldsgem}
    \end{subfigure}
    \hfill
    \begin{subfigure}[b]{0.3\textwidth}  
        \centering
        \includegraphics[width=\textwidth, height=5cm]{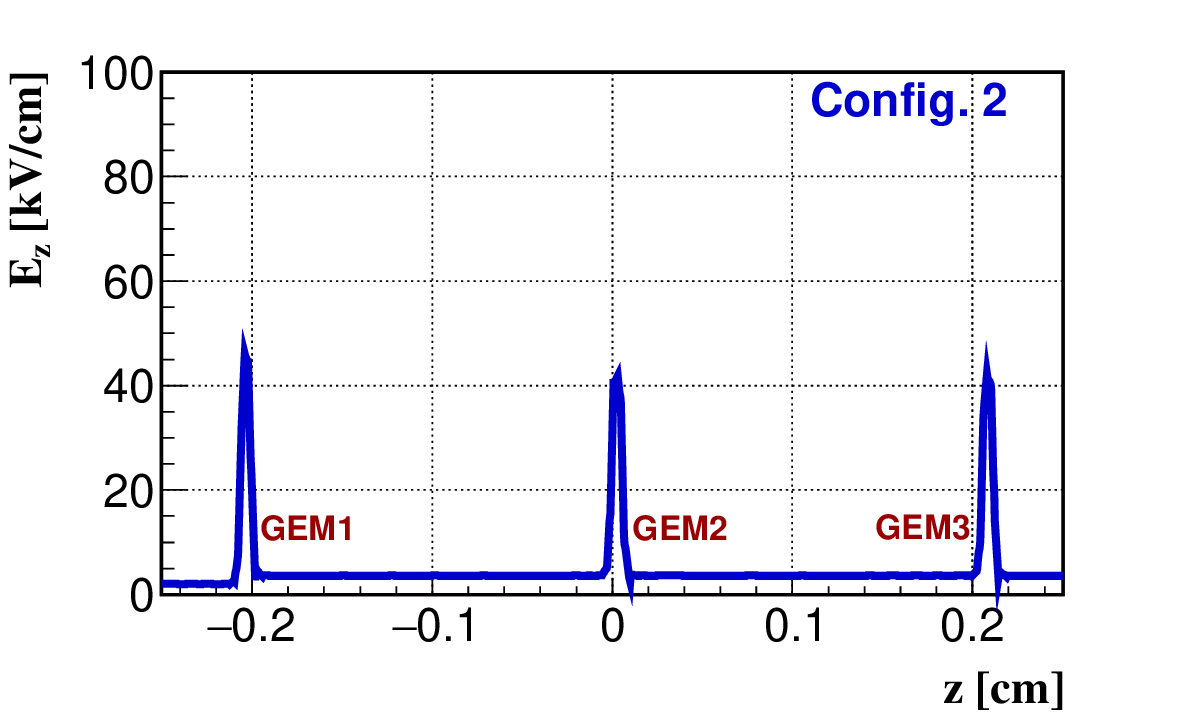}
        \caption{}
         \label{fig:fldfgem}
    \end{subfigure}
    \hfill
    \begin{subfigure}[b]{0.3\textwidth}  
        \centering
        \includegraphics[width=\textwidth, height=5cm]{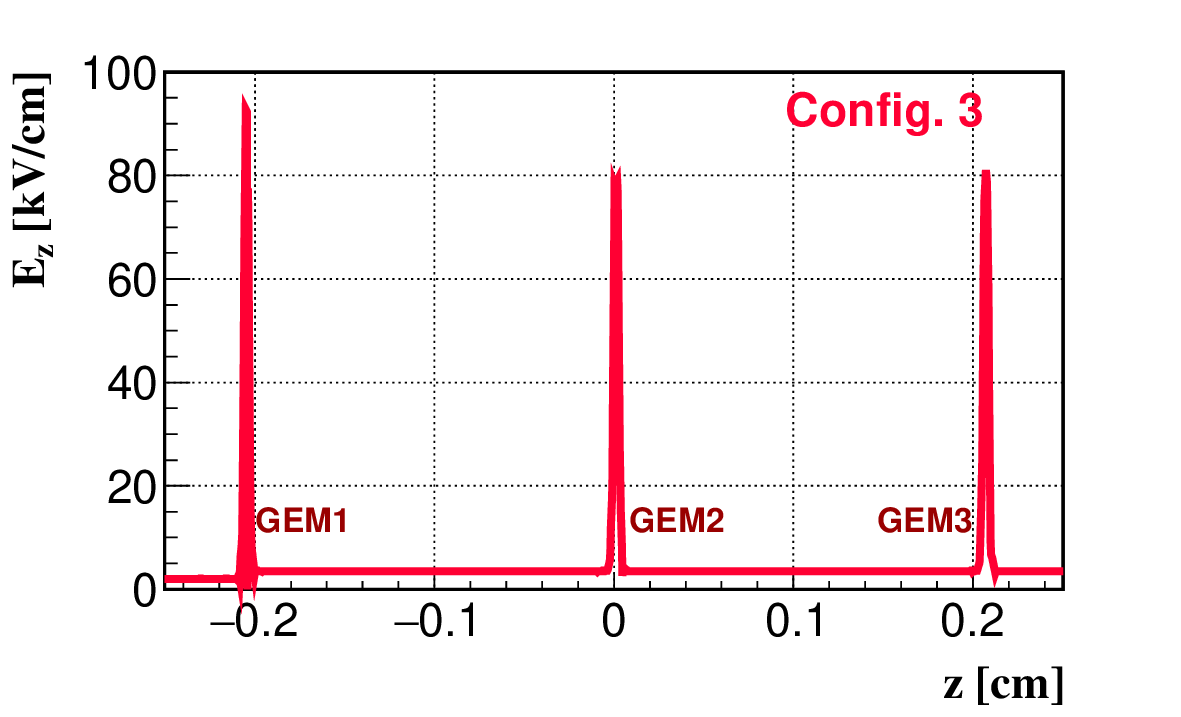}
        \caption{}
        \label{fig:fldftgem}
    \end{subfigure}

    \caption{Simulated peak electric field strength (E$_{z}$) inside the GEM holes across three GEM configurations : (a) Config. 1 (b) Config. 2 (c) Config. 3 at $\Delta V_{\mathrm{GEM}}$ of 300 V.}
    \label{fldgem}
    
\end{figure*}

\begin{figure*}[t!]
    \centering
    \begin{subfigure}[b]{0.3\textwidth} 
        \centering
        \includegraphics[width=\textwidth, height=5cm]{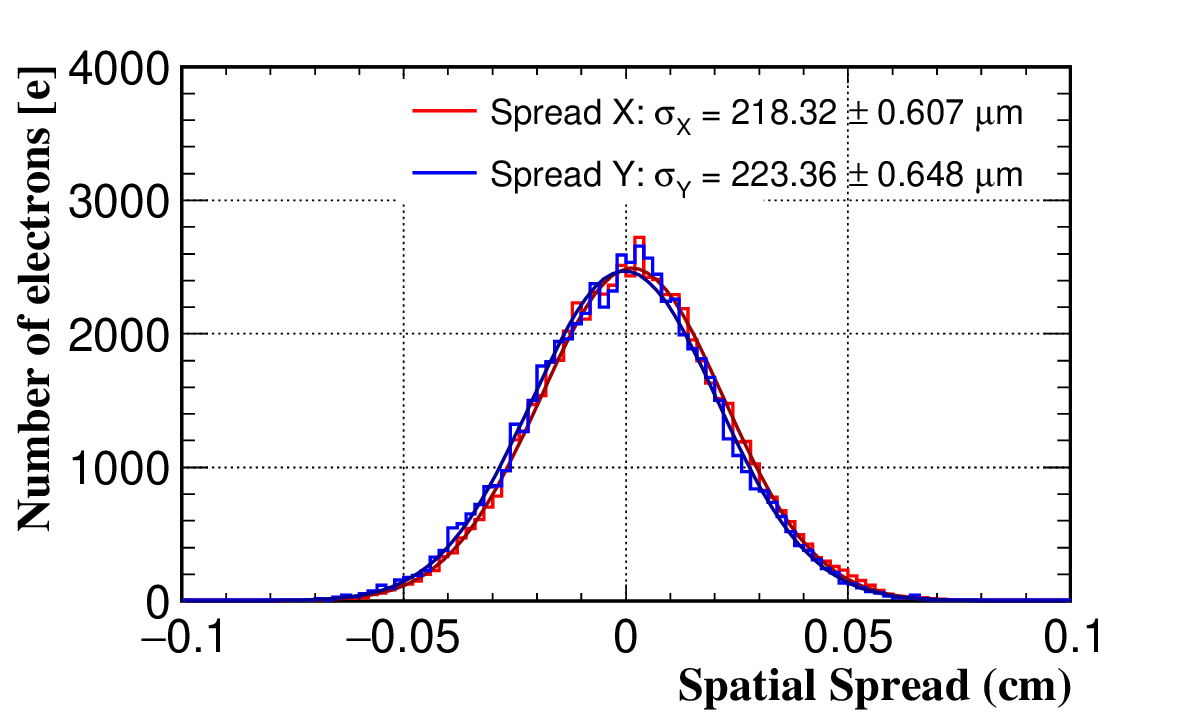}  
        \caption{}
        \label{fig:xysgem}
    \end{subfigure}
    \hfill
    \begin{subfigure}[b]{0.3\textwidth}  
        \centering
        \includegraphics[width=\textwidth, height=5cm]{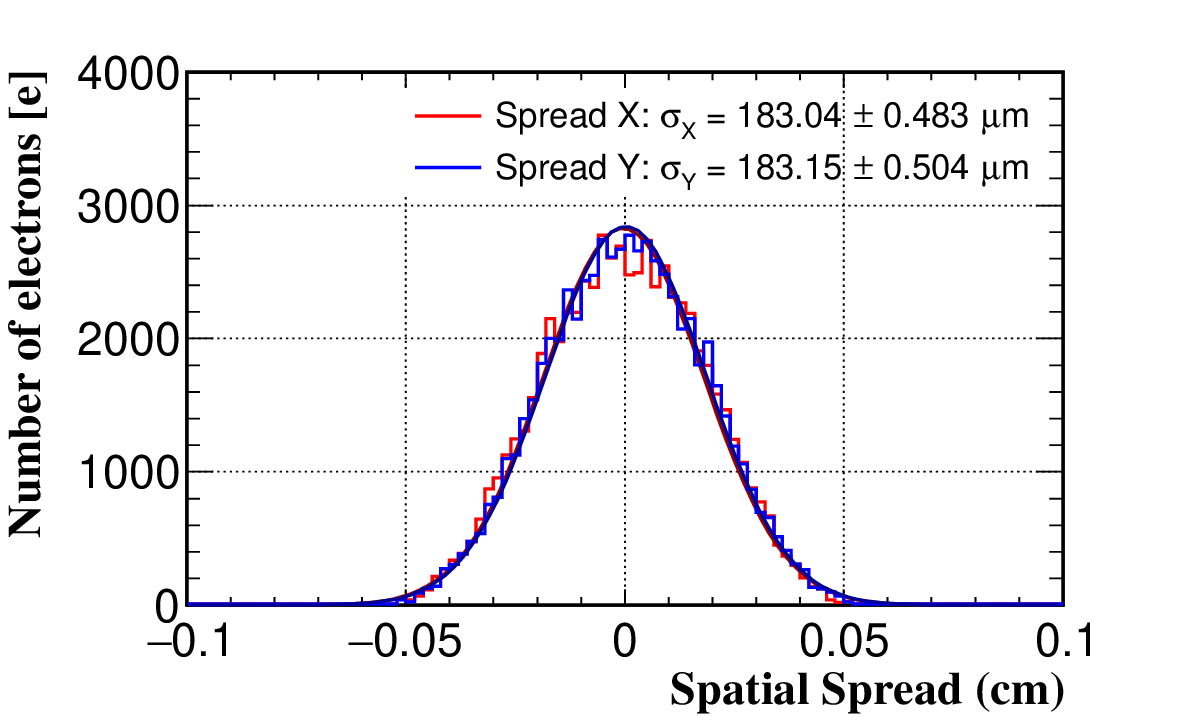}
        \caption{}
         \label{fig:xyfgem}
    \end{subfigure}
    \hfill
    \begin{subfigure}[b]{0.3\textwidth}  
        \centering
        \includegraphics[width=\textwidth, height=5cm]{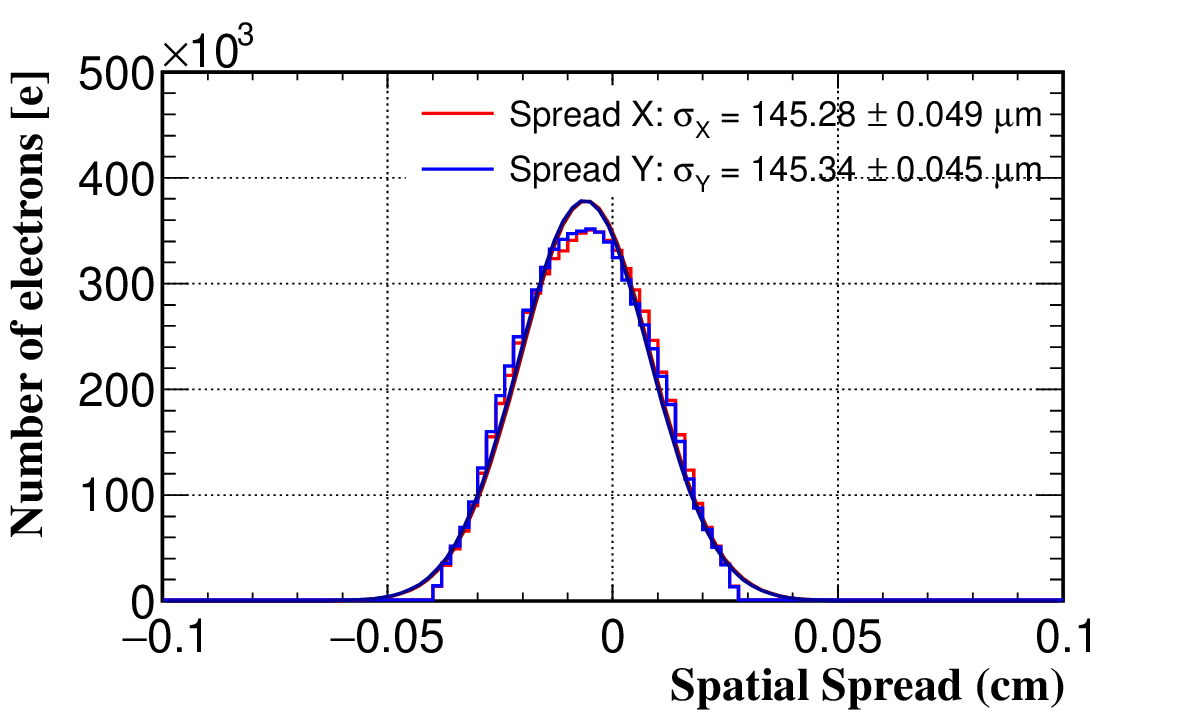}
        \caption{}
        \label{fig:xyftgem}
    \end{subfigure}

    \caption{Electron spread with Gaussian fits on the induction electrode for (a) Config. 1 (3 SGEMs), (b) Config. 2 (3 FGEMs), and (c) Config. 3 (3 FTGEMs), based on 1000 simulated events at $\Delta V_{\mathrm{GEM}}$ of 300 V. The standard deviations ($\sigma_{x,y}$) of the electron cloud for each configuration are also provided.}
    \label{fig:xygem}
    
\end{figure*}

Figs. \ref{fig:dlsgem} -- \ref{fig:dlftgem} show the drift lines of electrons at $\Delta V_{\mathrm{GEM}}$ of 300 V across Configs. 1, 2, and 3, respectively. The density of drift lines increases from Config. 1 to Config. 3, indicating the enhancement of avalanche effects due to a strong localized electric field within a narrow hole region. Moreover, the number of secondary electrons drifting out of GEM1 is higher, due to which more electrons participate in the avalanche in the subsequent GEM holes, resulting in increased gain.

Figs. \ref{fig:fldsgem} -- \ref{fig:fldftgem} shows the peak electric field strength (E$_\text{z}$) within the GEM holes across Configs. 1, 2, and 3 at a fixed $\Delta V_{\mathrm{GEM}}$ of 300 V. The reduced pitch GEM configuration, particularly Config. 3 shows a noticeably higher localized E$_\text{z}$ for the same GEM potential. E$_\text{z}$ in Configs. 1 and 2 range between 40--45 kV/cm, whereas Config. 3 can achieve nearly 100 kV/cm. The enhanced E$_\text{z}$ at low potential for Config. 3 indicates that this geometry should be operated at a lower potential.

The endpoints of the electrons produced during the avalanche events are tracked to measure electron losses and their transmission in the corresponding transfer regions. EUM1, EUM2, and EUM3 refer to electrons that are collected in the upper metal regions of the corresponding GEM structures. EKAP1, EKAP2, and EKAP3 indicate trapped electrons within the GEM holes. Collectively, EUM and EKAP represent the total electron loss by GEM's imperfect electron transparency. ETRAN1, ETRAN2, and ETRAN3 refer to electrons that are collected in Tran. 1, Tran. 2, and Tran. 3 regions, respectively. It is worth noting that, although the electrons may hit the lower metal surfaces in the stacked GEMs, the endpoint of avalanche electrons at the lower metal was estimated to be negligible. It may be due to the simulation tracking the endpoints after the avalanche effect has occurred. Therefore, the electrons drifting back after the avalanche and terminating at the lower GEM metals are likely negligible, owing to the directionality of the applied electric field. 

The difference between the reconstructed endpoint position of the avalanche electrons and the true initial position of the primary electrons is represented in a histogram \cite{Lan-Lan:2013dxa}. This histogram is then used to determine the intrinsic position resolution in terms of the standard deviation of the electron spreads (\(\sigma_{x}\), \(\sigma_{y}\)), as the distribution follows a Gaussian profile. Figs. \ref{fig:xysgem}, \ref{fig:xyfgem}, and \ref{fig:xyftgem} show the X and Y spreads of avalanche electrons across Config. 1, Config. 2, and Config. 3, respectively. These spreads are presented for $\Delta V_{\mathrm{GEM}}$ of 300 V. The distribution amplitude is greater for Config. 3 followed by Config. 2 in comparison to Config. 1, indicating enhanced electron multiplication resulting from reduced hole pitch. The peak-fitting Gaussian function  \cite{Lan-Lan:2013dxa} provides an effective description of the Gaussian profile for the simulated GEM configurations and is expressed as:
 
\begin{equation} \label{eq:gauss}
\begin{aligned}
y &= \left( \frac{A}{w \sqrt{\pi / 2}} \right)
     \exp\!\left( -2\,\frac{(x - x_c)^2}{w^2} \right), \\
      \sigma_{x,y} &\approx \frac{w}{2},
\end{aligned}
\end{equation}

where $w$ [$\mu$m] is the width of the Gaussian profile, $y$ [$\mu$m] denotes the function value, $A$ [$\mu$m] is the amplitude, $x$ [$\mu$m] is the independent variable, and  $x_c$ [$\mu$m] corresponds to the mean position. The resultant standard deviation of the electron spread can be quantified using:
\begin{equation}
    \sigma = \sqrt{\sigma_{x}^2 + \sigma_{y}^2}.
\end{equation}

It is evident from Fig. \ref{fig:xygem} that the reduced pitch GEM detectors (Configs. 2 and 3) exhibit a lower $\sigma$ value compared to the standard configuration (Config. 1). The reduction in $\sigma$ may be attributed to collimated electric field lines that constrain the transverse diffusion. A reduced intrinsic spread is generally expected to improve the overall position resolution of the detector \cite{Flothner:2024qev}. However, the final position resolution is also influenced by several additional factors, including the strip pitch \cite{Kudryavtsev:2017cbu}, the readout electronics \cite{Flothner:2024qev}, the magnetic field \cite{Alexeev:2019rng}, the multiple scattering \cite{Kudryavtsev:2020ohh}, and the geometrical alignment of readout strips \cite{Bressan:1998jj}. Broadly, the total position resolution ($\sigma_{\text{pos}}$) can be approximated as:
\begin{equation} \label{posresol}
\sigma_{\text{pos}}^{2}
= \sigma_{\text{0}}^{2}
+ \sigma_{\text{meas}}^{2},
\end{equation}
where $\sigma_{\text{0}}$ depends on the transverse diffusion of electron clouds and follows the transport dynamics. It is related to the transverse diffusion coefficient ($D_{T}$) by the following equation \cite{Trindade:2023}:
\begin{equation} \label{eqndiffusion}
\sigma_{\text{0}} = \sqrt{\frac{4 \, s \, D_T}{W}}, \quad \text{where} \quad D_x = D_y = 2 D_T, and \quad W = \frac{eE}{\varepsilon_{KT}}.
\end{equation}

Here, $s$ is the drift length, $\varepsilon_{KT}$ is the characteristic energy associated with transverse diffusion, and $E$ is the applied electric field. $\sigma_{\text{0}}$ helps determine uncertainty in the intrinsic spatial resolution of the electron cloud at the readout. Its measurement is independent of readout geometry and electronics. 

On the other hand, $\sigma_{\text{meas}}$, which is not accounted for in the simulation, represents the contribution to the absolute position resolution from readout strips/pads and associated electronics. It primarily depends on the $D_{T}$, the pitch of the readout strips, the signal-to-noise ratio (SNR), the signal thresholds, and the reconstruction algorithm. For a single electrode, $\sigma_{\text{meas}}$ is primarily dominated by a factor of pitch/$\sqrt{12}$. In contrast, for a multi-strip readout, it is largely governed by the standard deviation of residuals obtained using the Centre of Gravity (COG) method. For an effective COG methodology, electron diffusion plays a vital role. If the diffusion is larger, the electron clouds spread over many strips, which broadens the cluster, degrades the centroid precision, and reduces the SNR. Conversely, if the diffusion is too small, the electron cloud is collected entirely on the single strip, limiting the resolution to the pitch size. Therefore, the optimal level of diffusion, such that they can diffuse over two to three strips, is desirable for achieving good position resolution. Based on the aforementioned factors, the simulation results are discussed in the subsequent sections.

\section{Comparison between simulation and experimental gain\label{correlation}}

\begin{figure}[htbp]
\centering
\includegraphics[width=.49\textwidth,height=5cm]{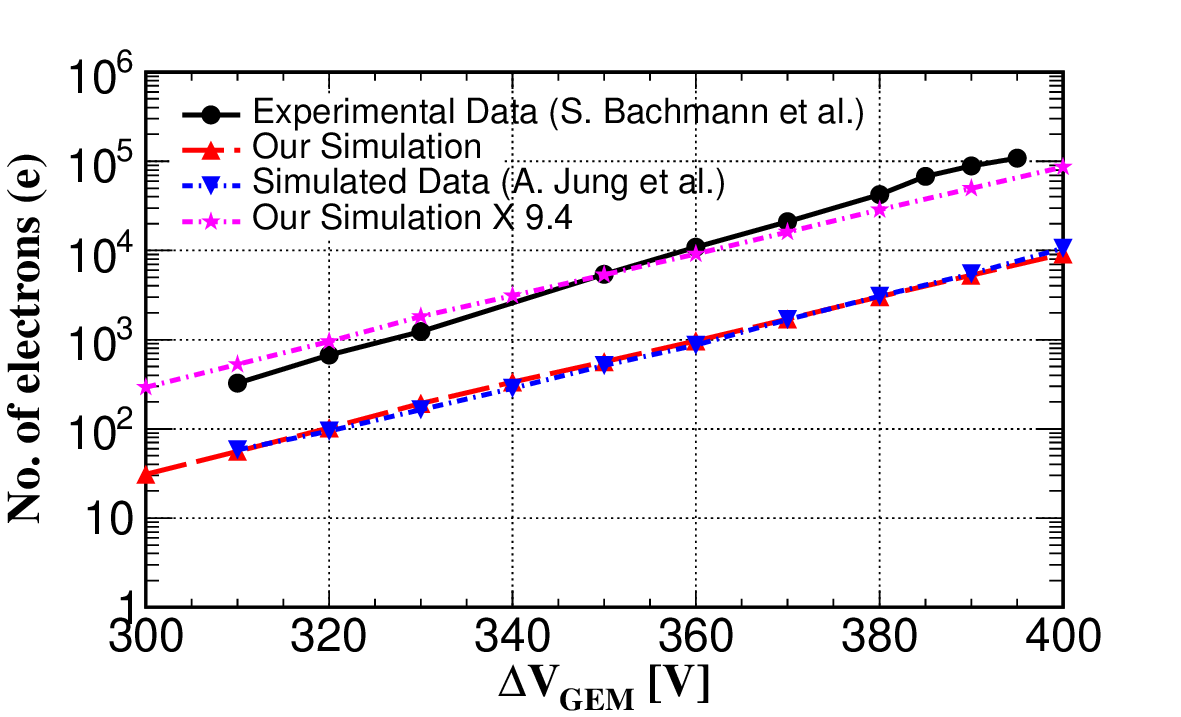}
\caption{Number of electrons collected in Config. 1 (3 SGEMs) as a function of $\Delta V_{\mathrm{GEM}}$. The experimental data are sourced from Ref. \cite{Bachmann:1999xc}, and the simulated data are from Ref. \cite{Jung:2021cvz}.\label{fig4}}
\end{figure}

Fig. \ref{fig4} shows the number of collected electrons in Config. 1 across Tran. 3 region as a function of $\Delta V_{\mathrm{GEM}}$. We compare these electron collections with existing experimental data \cite{Bachmann:1999xc} and simulated results \cite{Jung:2021cvz} for validation and analysis. The Penning transfer ratio for incorporating the Penning effect is 0.57. Penning effect refers to the process in which an excited atom transfers part of its energy to a nearby quenching gas molecule, thereby enhancing the gain. The figure shows that the simulation exhibits a trend similar to the experimental data \cite{Bachmann:1999xc} regarding the increase in gain with increasing $\Delta V_{\mathrm{GEM}}$. Moreover, our simulated data shows good agreement with the previous simulated gain results \cite{Jung:2021cvz}. To provide an efficient comparison of gain with experimental measurements, we introduce an offset/multiplication factor ($M$) of 9.40. This value has been obtained using a log-based least-squares method and is given by the following equation:

\begin{equation} \label{eq:offset}
M = \exp\left( \frac{1}{n} \sum_{i=1}^{n} \left[ \ln(Gain)_{\text{exp}, i} - \ln(Gain)_{\text{sim}, i} \right] \right).
\end{equation}

Here, $i$ denotes the number of entries, $Gain$ indicates collected electrons in the induction region, $exp$ refers to the experimental data, and $sim$ corresponds to the simulated data.

Despite the reliable slope and overall functional dependence, the noticeable difference observed between the simulated and experimental data \cite{Bachmann:1999xc} reflects the characteristic normalization difference that is commonly reported in GEM simulations. Notably, increasing the GEM foils (multilayer GEM) increases the offset. For instance, the offset factor reported for a single GEM ranges from 1.3 to 2.25 \cite{Gupta:2025yrv, Jung:2021cvz, Bencivenni:2002iw, Sharma:1999nf, Janssens2019}. While in a triple GEM configuration, the deviation becomes much more pronounced \cite{Jung:2021cvz}. This large deviation in the triple GEM detector is attributed to the cumulative effect of intrinsic limitations of the model framework (e.g., the absence of a photon feedback mechanism, space charge interactions, imperfections in solving field maps, etc.) as discussed in Section \ref{section2}. These effects become more significant in multi-GEM configurations. Since the causes of this deviation are still being explored \cite{Sahin:2014haa, Rout:2021mbn, Azevedo:2016mer}, and we aim to provide a qualitative interpretation of the results, we have not included the offset factor in the subsequent analysis. 

In the below sections (Sections \ref{vgem}--\ref{gasconcentration}), the contribution from individual parameters is assessed by varying them independently, while keeping others constant. This approach helps understand the influence of each parameter on overall transport dynamics.

\section{Impact of GEM potential ($\Delta V_{\mathrm{GEM}}$ \label{vgem})}

\begin{figure*}[htbp]
\centering
\includegraphics[width=.99\textwidth,height=6.5cm]{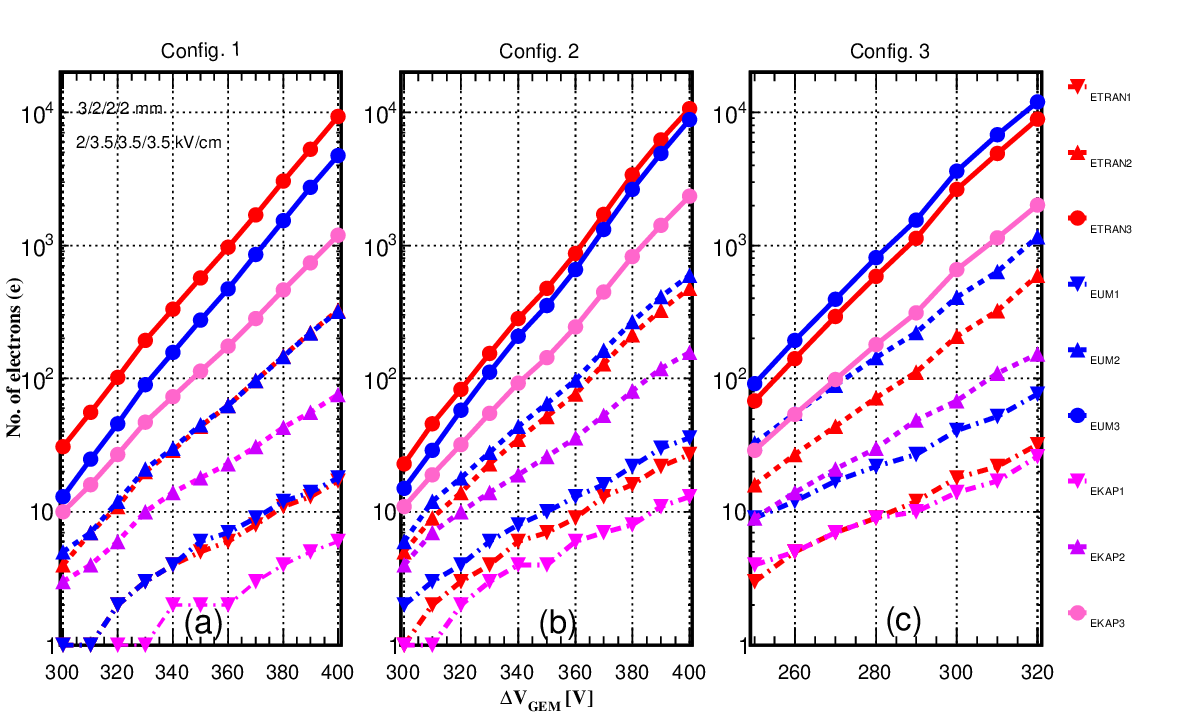}
\caption{Number of electrons transferred to the transfer regions in Configs. 1, 2, and 3 as a function of $\Delta V_{\mathrm{GEM}}$. The drift electric field is set to 2 kV/cm, and the transfer electric fields are maintained at 3.5 kV/cm.}
\label{fig5}
\end{figure*}

To quantify the impact of $\Delta V_{\mathrm{GEM}}$ on electron transport, we evaluate the number of electrons collected in each region, as shown in Fig. \ref{fig5}. With increasing $\Delta V_{\mathrm{GEM}}$, the absolute number of transferred electrons also increases, indicating higher avalanche multiplication at increased $\Delta V_{\mathrm{GEM}}$. The cascade pattern ($\mathrm{ETRAN3} > \mathrm{ETRAN2} > \mathrm{ETRAN1}$
) is consistently observed across the simulated GEM configurations. Configs. 1 and 2 show comparable electron collection in the transfer regions, whereas Config. 3 exhibits relatively higher electron transmission, likely due to its geometric effect and enhanced localized electric field strength. Moreover, the increased hole density allows a greater number of electrons to participate in the avalanche process \cite{Flothner:2024qev}. The observed cascade effect is consistent with the baseline finding.

The electron losses at the upper GEM metals (EUM1--EUM3) follow the same behavior as that of transferred electrons. This metallic absorption behavior likely occurs due to enhanced scattering of avalanche electrons \cite{Sauli:2016eeu} arising from diverging field lines that deviate the outgoing electrons towards the metal surfaces. In line with the cascading effect, the later GEM metals collect more electrons ($\mathrm{EUM3} > \mathrm{EUM2} > \mathrm{EUM1}$). Configurations with smaller pitch sizes show relatively higher electron loss. For instance, in Config. 3, the metallic absorption is consistently higher than the electron transmission. This behavior likely results from the higher ratio of metal to Kapton area, providing a greater surface for electron absorption. 

Electron trapping within the Kapton (EKAP1--EKAP3) similarly rises with increasing $\Delta V_{\mathrm{GEM}}$. This increase is attributed to the enhanced scattering of energetic secondary electrons towards the Kapton wall and formation of trapping centres due to space charge buildup within the holes \cite{Alfonsi:2012msp, Bouhali:2021vhg}. As seen in other channels, the highest trapping occurs in the last GEM ($\mathrm{EKAP3} > \mathrm{EKAP2} > \mathrm{EKAP1}$). Additionally, Config. 3 consistently shows the highest electron trapping across all GEM stages, likely due to a decrease in electron transparency, caused by the reduction of hole diameter. In contrast, Config. 1 shows the minimal trapping loss. The electrons lost due to trapping are less significant than those lost through absorption across all simulated GEM configurations.

\begin{table*}[h!]
\centering
\scriptsize
\caption{Electron losses and their corresponding weightages across different configurations. ETOTAL represents the total number of avalanche electrons. EKAP1, EKAP2, and EKAP3 denote the number of electrons lost within the respective GEM holes, while EUM1, EUM2, and EUM3 correspond to the electrons lost at the respective upper metal surfaces.}
\begin{tabular}{l|c|c|c|c|c|c|c}
\hline
\textbf{Configuration} & \textbf{ETOTAL} [e] & \textbf{EKAP1} [e] & \textbf{EKAP2} [e] & \textbf{EKAP3} [e] & \textbf{EUM1} [e] & \textbf{EUM2} [e] & \textbf{EUM3} [e] \\
\hline
Config. 1 [$\Delta V_{\mathrm{GEM}}$=360 V] & 1786 & 2 [0.11\%] & 23 [1.29\%]  & 175 [9.80\%] & 7 [0.39\%]  & 63 [3.53\%] & 473 [26.48\%]  \\
Config. 2 [$\Delta V_{\mathrm{GEM}}$=360 V] & 2025 & 6 [0.30\%] & 36 [1.77\%]  & 245 [12.10\%] & 13 [0.64\%] & 98 [4.84\%] & 660 [32.59\%]  \\
Config. 3 [$\Delta V_{\mathrm{GEM}}$=290 V] & 3431 & 10 [0.29\%]  & 46 [1.34\%]  & 312 [9.09\%]  & 27 [0.79\%] & 222 [6.47\%] & 1551 [45.20\%] \\
\hline
\end{tabular}
\label{tab:tab2}
\end{table*}

\begin{table}[h!]
\centering
\scriptsize
\caption{Electron transfer and their corresponding weightages across different configurations. ETRAN1, ETRAN2, and ETRAN3 represent the number of electrons collected in the Transfer 1, Transfer 2, and Induction regions, respectively.}
\begin{tabular}{l|c|c|c}
\hline
\textbf{Configuration} & \textbf{ETRAN1} (e) & \textbf{ETRAN2} (e) & \textbf{ETRAN3} (e) \\
\hline
Config. 1 [$\Delta V_{\mathrm{GEM}}$=360 V] & 6 [0.33\%]    & 63 [3.53\%]  &  966 [54.08\%]\\
Config. 2 [$\Delta V_{\mathrm{GEM}}$=360 V] &  9 [0.44\%]  & 77 [3.80\%]   &  874 [43.16\%]\\
Config. 3 [$\Delta V_{\mathrm{GEM}}$=290 V] & 12 [0.34\%]   & 112 [3.26\%]  & 1132 [32.99\%] \\
\hline
\end{tabular}
\label{tab:tab3}
\end{table}

\begin{table*}[ht]
\caption{Values of the parameters of position resolution at $\Delta \text{V}_\text{{GEM}}$= 300 V.}\label{tab:table4}

\centering
\scriptsize
\begin{tabular}{c|c|c|c|c|c|c}

\hline
\textbf{GEM Type} & $\sigma_{x}$ [µm] & \textbf{Reduction (\%)} & $\sigma_{y}$ [µm] & \textbf{Reduction (\%)} & $\sigma$ [µm] & \textbf{Reduction (\%)} \\ 
\hline
Config. 1 &  207.46$\pm$0.573 & -- &208.93$\pm$0.584 &--& 294.43 $\pm$0.580&-- \\ 
Config. 2 &  183.04$\pm$0.483 & 11.77 & 183.15$\pm$0.504 & 12.34 & 258.94 $\pm$0.494 & 12.05  \\ 
Config. 3 & 145.28$\pm$0.049 & 29.97 & 145.34$\pm$0.045 & 30.43 & 205.50 $\pm$0.047 & 30.32\\ 

\hline

\end{tabular}
\end{table*}

Table \ref{tab:tab2} presents the number of electrons lost through metallic absorption and trapping, along with their percentage contribution. Table \ref{tab:tab3} lists the corresponding electron transmission to the transfer regions (ETRAN3). The data are presented at $\Delta V_{\mathrm{GEM}}$=360 V for Configs. 1 and 2, and at 290 V for Config. 3, adopted to ensure comparable and sufficient electron transfer ($\sim$10$^{3}$) in the induction region. Although Config. 3 achieves higher electron collection at low $\Delta V_{\mathrm{GEM}}$, a larger fraction of avalanche electrons is lost within the GEM system, leading to reduced transparency.

\begin{figure*}[h!]
    \centering
    \begin{subfigure}[b]{0.48\textwidth} 
        \centering
        \includegraphics[width=\textwidth, height=5cm]{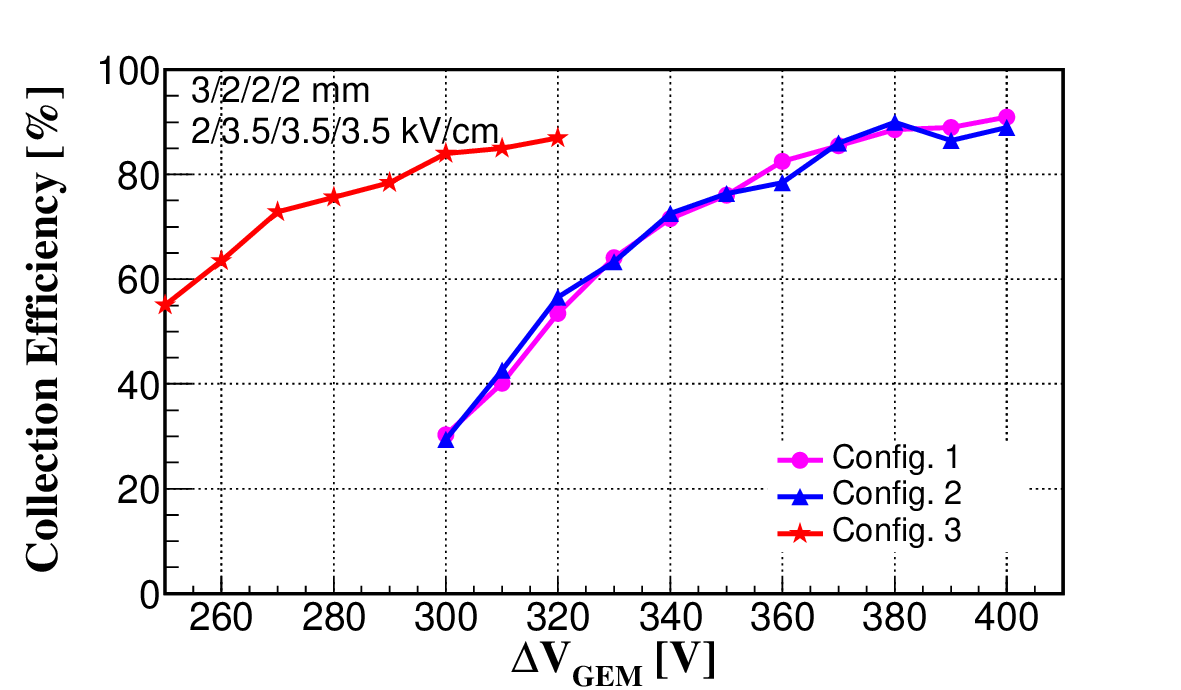}  
        \caption{}
        \label{fig:cevgem}
    \end{subfigure}
    \hfill
    \begin{subfigure}[b]{0.48\textwidth}  
        \centering
        \includegraphics[width=\textwidth, height=5cm]{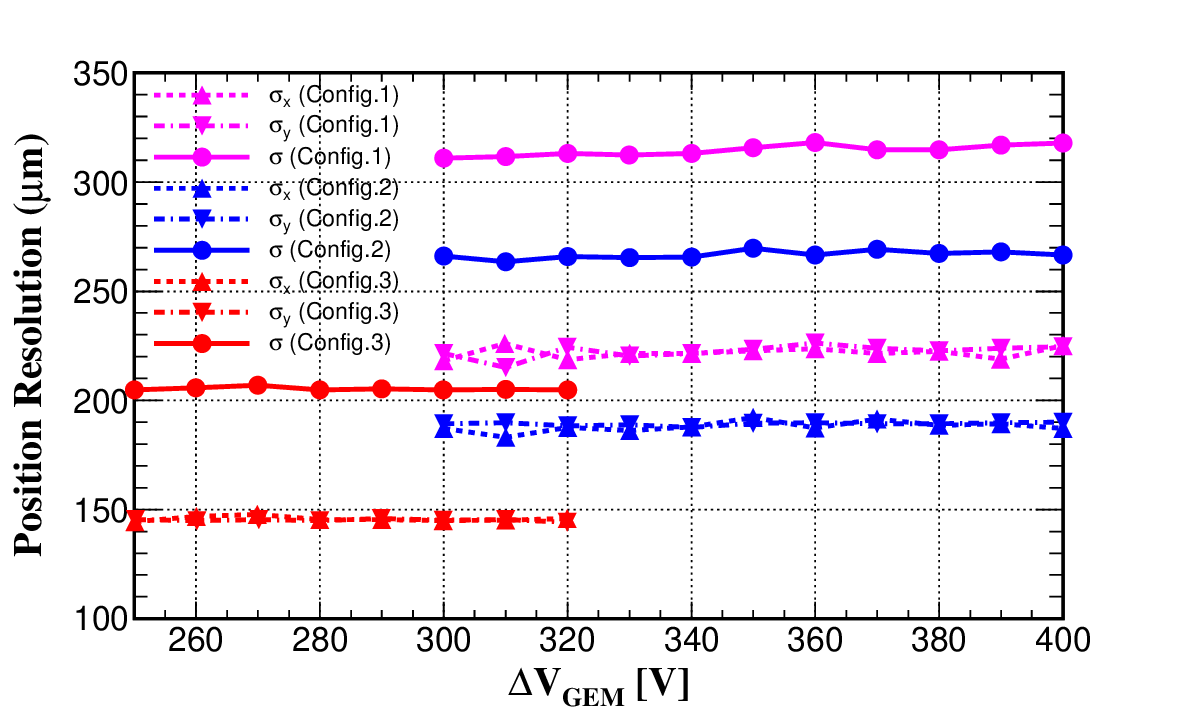}
        \caption{}
        \label{fig:resolution}
    \end{subfigure}

    \caption{(a) Collection efficiency as a function of $\Delta V_{\mathrm{GEM}}$ for three distinct GEM configurations. (b) Variation of position resolution ($\sigma_x$, $\sigma_y$, $\sigma$) with $\Delta V_{\mathrm{GEM}}$. The drift and transfer electric fields are set to 2 kV/cm and 3.5 kV/cm, respectively.}
    \label{fig:6}
\end{figure*}

Fig. \ref{fig:cevgem} shows the variation of collection efficiency (CE) as a function of $\Delta V_{\mathrm{GEM}}$ across the simulated GEM configurations. It is evident from the figure that the CE improves with increasing $\Delta V_{\mathrm{GEM}}$. Configs. 1 and 2 exhibit nearly identical CE behavior with increasing $\Delta V_{\mathrm{GEM}}$, whereas Config. 3 can attain higher CE even at low $\Delta V_{\mathrm{GEM}}$. The improvement of CE with increasing $\Delta V_{\mathrm{GEM}}$ can be attributed to the effective pulling of electrons within the GEM holes, reducing their loss to the metal surfaces. The enhanced CE in Config. 3 even at low $\Delta V_{\mathrm{GEM}}$ may be attributed to enhanced localized electric field.

Fig. \ref{fig:resolution} shows the variation of position resolutions ($\sigma_x$, $\sigma_y$, $\sigma$) as a function of $\Delta V_{\mathrm{GEM}}$ for the simulated GEM configurations. $\sigma_x$, $\sigma_y$, and the overall $\sigma$ remain nearly constant with increasing $\Delta V_{\mathrm{GEM}}$ across all three configurations, indicating negligible impact of $\Delta V_{\mathrm{GEM}}$ on electron sampling. The figure also shows that reducing the pitch size enables finer sampling of the electron cloud, leading to reduced $\sigma_x$, $\sigma_y$, and $\sigma$ values. The position resolution of the modeled GEM configurations at $\Delta V_{\mathrm{GEM}}$ of 300 V and their percentage ($\%$) reduction compared to Config. 1 are summarized in Table \ref{tab:table4}.

Based on microscopic tracking across the simulated GEM configurations, we identify three key baseline transport characteristics: 
 
 \begin{enumerate}
     \item Cascade effect- The final GEM (GEM3) consistently exhibits higher electron loss and transmission, followed by GEM2 and GEM1.
     \item Loss mechanism- In almost all simulated cases, the metallic absorption (EUM) remains the dominant loss mechanism, while trapping loss (EKAP) serves as the secondary contributor.
     \item Geometric effect- Reducing the pitch size (Config. 1 $\to$ Config. 3) reduces the intrinsic spread ($\sigma$) of the electron cloud.
     
 \end{enumerate}

\begin{figure*}[!t]
\centering
\includegraphics[width=.99\textwidth,height=6.5cm]{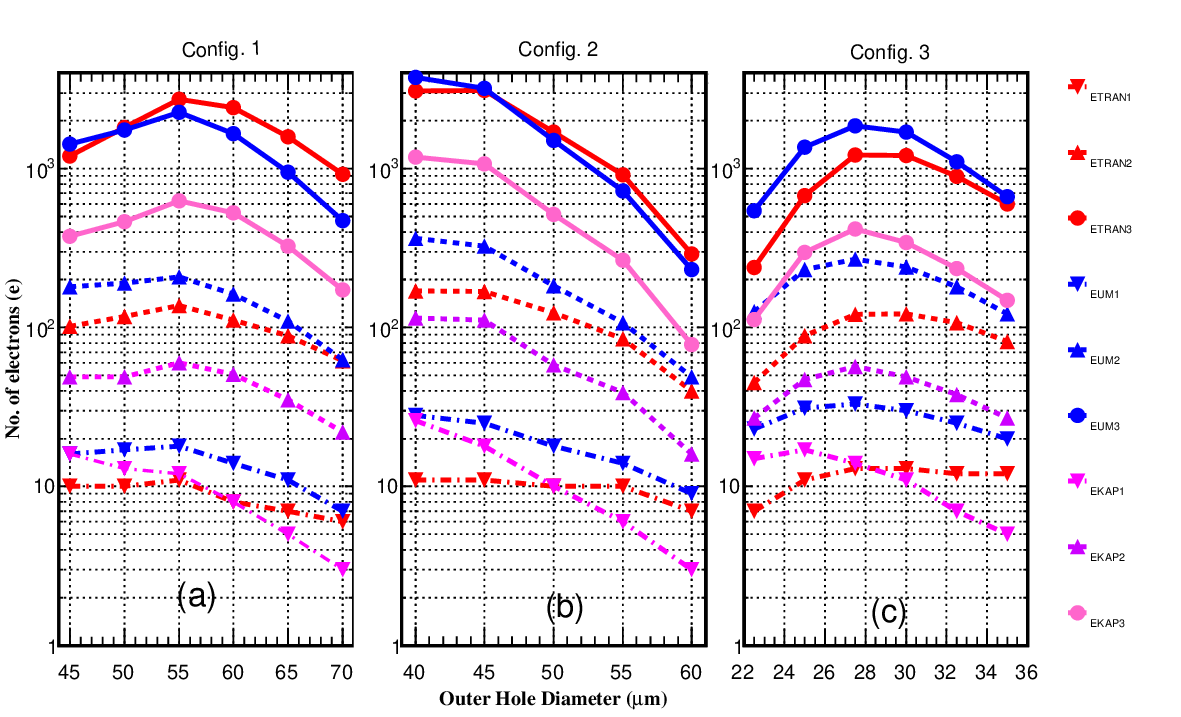}
\caption{Electron collections across Transfer, Upper Metal, and Kapton regions as a function of outer hole diameter for (a) Config. 1 at $\Delta V_{\mathrm{GEM}}$ = 360 V, (b) Config. 2 at $\Delta V_{\mathrm{GEM}}$ = 360 V, and (c) Config. 3 at$\Delta V_{\mathrm{GEM}}$ = 290 V. \label{fig11}}
\end{figure*}

\begin{figure*}[h!]
    \centering
    \begin{subfigure}[b]{0.48\textwidth} 
        \centering
        \includegraphics[width=\textwidth, height=5cm]{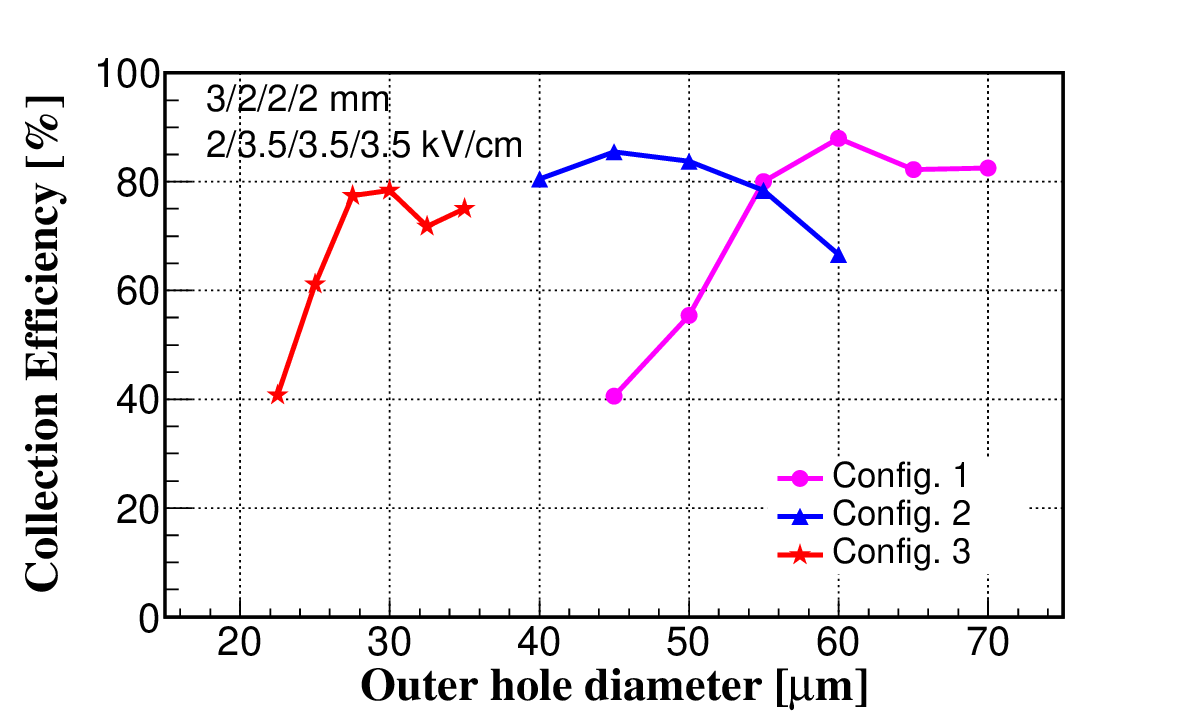}  
        \caption{}
        \label{fig:8a}
    \end{subfigure}
    \hfill
    \begin{subfigure}[b]{0.48\textwidth}  
        \centering
        \includegraphics[width=\textwidth, height=5cm]{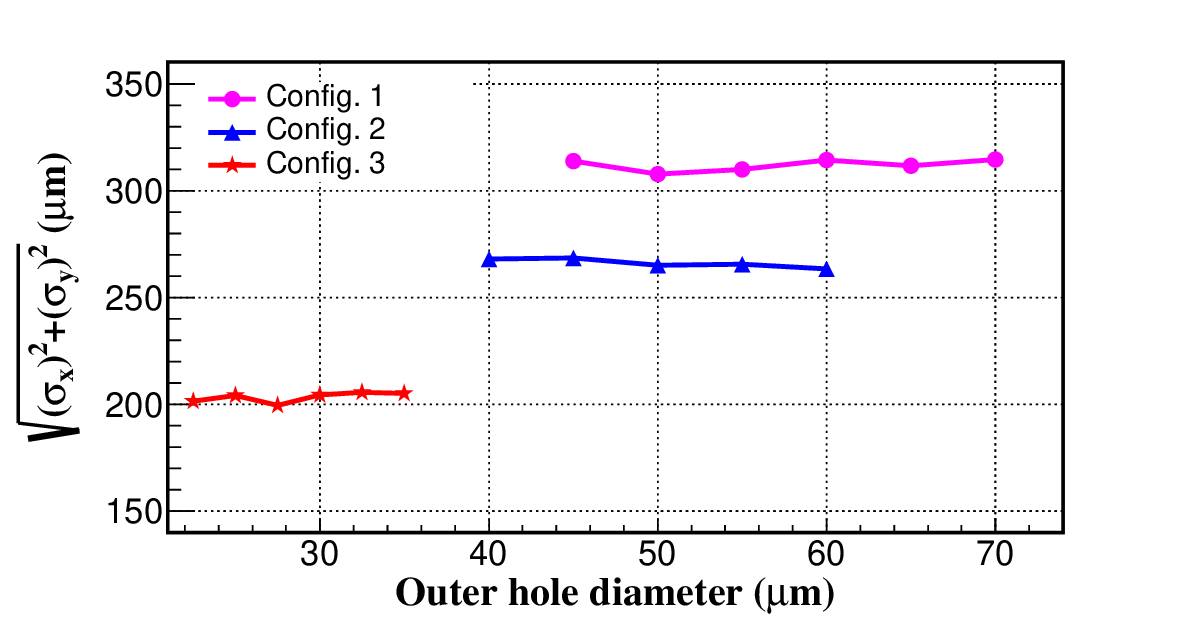}
        \caption{}
        \label{fig:8b}
    \end{subfigure}

    \caption{(a) Collection efficiency as a function of outer hole diameter for three distinct GEM configurations. (b) Variation of  \(\sigma = \sqrt{\sigma_{x}^2 + \sigma_{y}^2}\) with outer hole diameter. The drift and transfer electric fields are set to 2 kV/cm and 3.5 kV/cm, respectively.}
    \label{fig:6}
\end{figure*}

\section{Impact of outer hole diameter \label{outerholediameter}}

Fig.~\ref{fig11} shows the absolute number of collected electrons in the transfer regions, on the GEM electrodes, and inside the GEM holes as a function of outer hole diameter. The absolute number of transferred electrons follows a consistent trend across Configs. 1 and 3: increasing the outer hole diameter results in an initial rise followed by a gradual decline after reaching a specific diameter. Configs. 1 and 3 exhibit maximum electron collection at a hole diameter of
55 $\mu$m and 27.5 $\mu$m, respectively. While the maximum electron transfer in Config. 2 is observed at an initial hole diameter of 40 $\mu$m. The initial rise in electron collection in Configs. 1 and 3 are attributed to an increase in aperture for incoming electrons, whereas the reduction after a specific value suggests distortion may arise due to the distorted electric field lines. The electron collection in the Tran. 3 region follows a similar trend to that reported in the article for a single GEM \cite{Das:2015tpa, Mondal:2024ogs}. 

From the same figure, it is evident that the maximum electron absorption on the upper GEM electrodes also follows a consistent trend as that of the transferred electrons. In Config.~3, electron loss due to absorption becomes dominant over the transferred electrons across Tran.~3 region as the hole diameter increases, whereas in Config.~2, this dominance is observed at a hole diameter of 40~$\mu$m. The initial increase in metallic absorption may be attributed to the larger hole diameter that increases the divergence of the field lines. However, further increasing the outer hole diameter may enhance electron extraction, thereby decreasing metal absorption.

The hole diameters corresponding to maximum electron trapping also coincide with those showing maximum electron transfer and absorption, suggesting a common influence of hole geometry and electric field distribution. Consistent with baseline findings,  electron loss by trapping is lower than that by metallic absorption. The combined effect of the hole dimension and the surrounding electric field lines likely governs this behavior. As the diameter of the outer holes increases, a greater number of electrons are channeled into the GEM holes, leading to an enhanced avalanche process. Although the outer hole diameter widens, the higher number of avalanche electrons increases the likelihood of trapping. Moreover, a larger hole diameter, due to its possible impact on field distortions at hole edges, may further contribute to electron trapping. However, beyond a specific diameter, the enhanced extraction efficiency provided by the wider exit surface may reduce electron confinement, resulting in reduced electron trapping.

Fig.~\ref{fig:8a} shows the variation of collection efficiency (CE) as a function of outer hole diameter for the simulated GEM configurations. It is observed that each configuration exhibits a maximum CE at a particular dimension of outer hole diameter. At this dimension, enhanced electrofocusing of the field lines into the GEM holes likely improves electron collection, resulting in higher CE. The corresponding optimal outer hole diameter for each configuration ensures maximum electron transfer in the subsequent stages, as evidenced from Fig.~\ref{fig11}.

Fig.~\ref{fig:8b} shows that the $\sigma$ values remain nearly constant as the outer hole diameter increases across all configurations. The weak dependence of the $\sigma$ value on hole size is particularly interesting, as a wider outer hole diameter would intuitively be expected to produce a broader electron cloud on the induction plane. However, this trend is not observed, indicating that electron spreading primarily relies on the transverse diffusion process and the electric field distribution in the transfer regions. The reduction in $\sigma$ with decreasing pitch size is consistent with the third baseline finding.

 \begin{figure*}[t!]
\centering
\includegraphics[width=.99\textwidth,height=6.5cm]{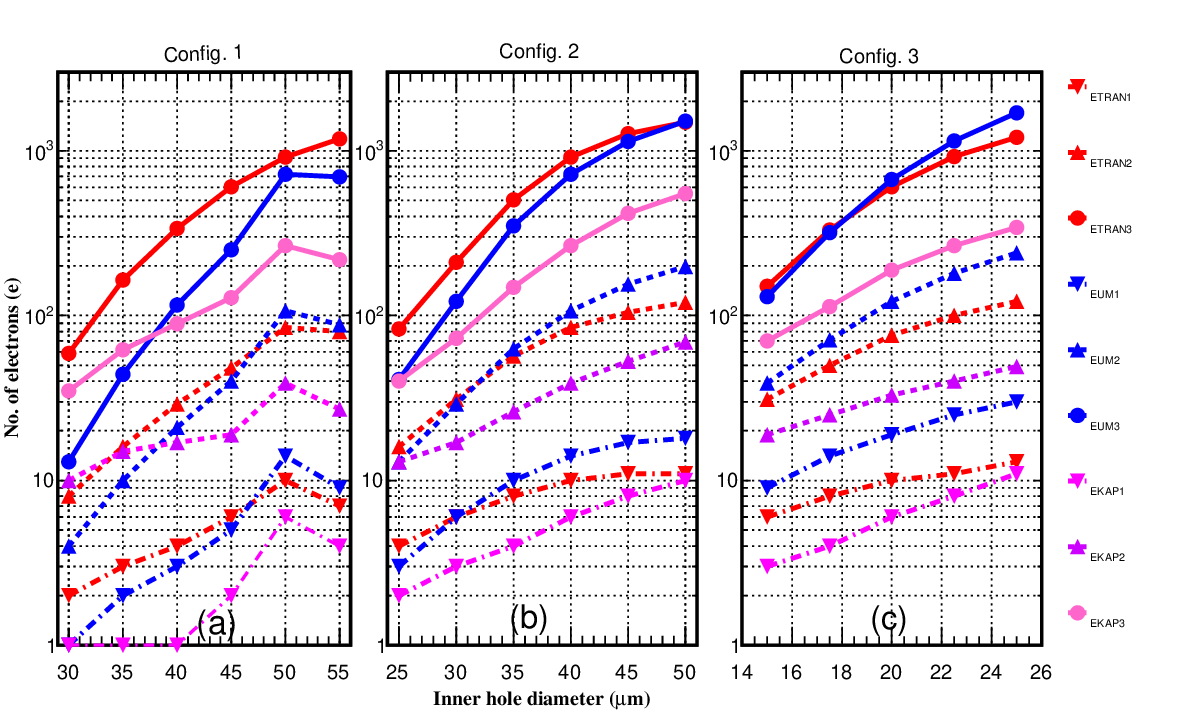}
\caption{Electron collections across Transfer, Upper Metal, and Kapton regions as a function of inner hole diameter for (a) Config. 1 at $\Delta V_{\mathrm{GEM}}$ = 360 V, (b) Config. 2 at $\Delta V_{\mathrm{GEM}}$ = 360 V, and (c) Config. 3 at$\Delta V_{\mathrm{GEM}}$ = 290 V. \label{fig13}}
\end{figure*}

\begin{figure*}[h!]
    \centering
    \begin{subfigure}[b]{0.48\textwidth} 
        \centering
        \includegraphics[width=\textwidth, height=5cm]{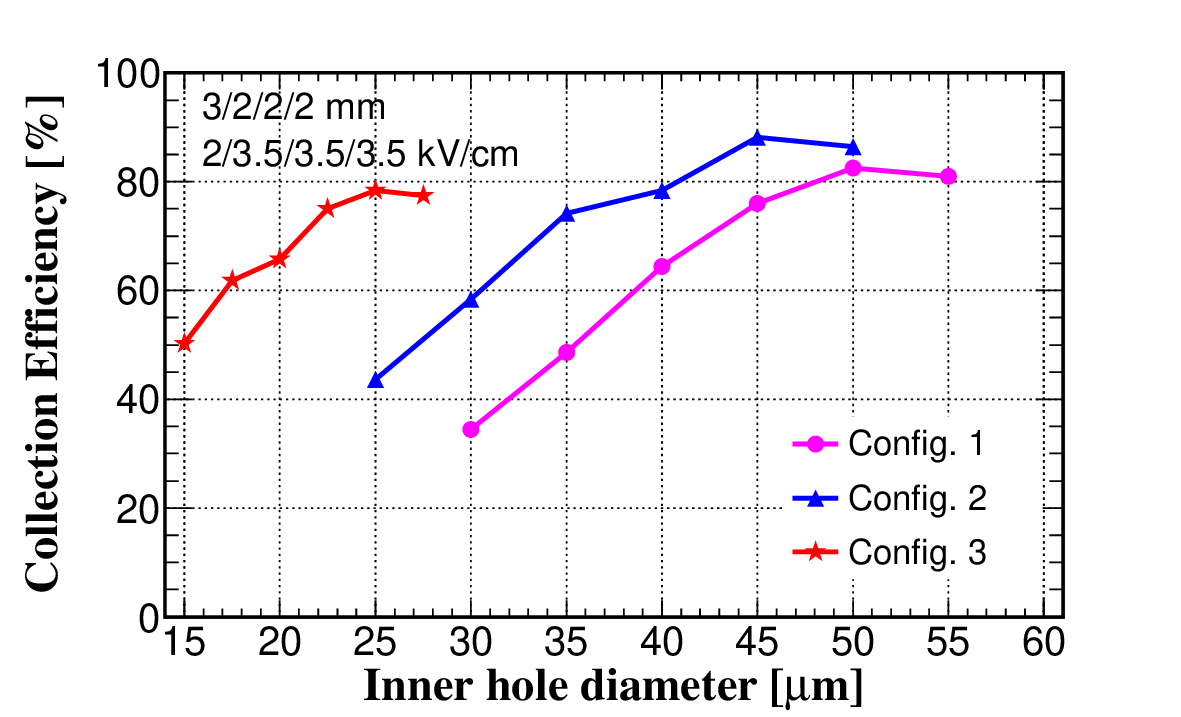}  
        \caption{}
        \label{fig:idce}
    \end{subfigure}
    \hfill
    \begin{subfigure}[b]{0.48\textwidth}  
        \centering
        \includegraphics[width=\textwidth, height=5cm]{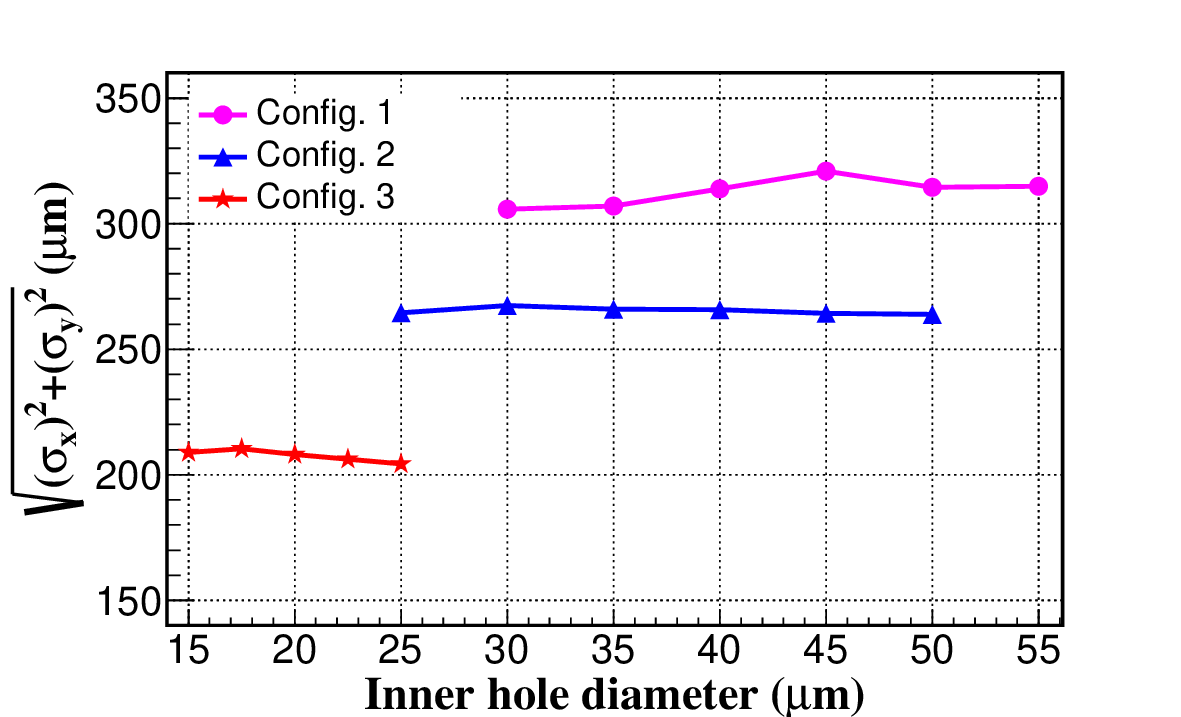}
        \caption{}
        \label{fig:idr}
    \end{subfigure}

    \caption{(a) Collection efficiency as a function of inner hole diameter for three distinct GEM configurations. (b) Variation of  \(\sigma = \sqrt{\sigma_{x}^2 + \sigma_{y}^2}\) with inner hole diameter. The drift and transfer electric fields are set to 2 kV/cm and 3.5 kV/cm, respectively.}
    \label{fig:10}
\end{figure*}

\section{Impact of inner hole diameter \label{innerholediameter}}

Fig. \ref{fig13} shows the absolute number of collected electrons in the transfer regions, on the GEM electrodes, and inside the GEM holes as a function of inner hole diameter. It is observed that the number of transferred electrons to the subsequent transfer regions increases as the inner hole diameter increases across all GEM configurations. This trend suggests that an increase in inner hole diameter expands the localized avalanche regions, thereby enhancing avalanche multiplication. As a result, more electrons are collected in the transfer regions. The increase in electron collection with increasing inner hole diameter, particularly in the induction region, is consistent with prior observation \cite{Das:2015tpa}.

The figure also indicates that electron absorption at the upper GEM metal surfaces becomes more significant as the inner hole diameter increases. As the interaction region gets broader, the electric field lines may become less localized at the hole edges. Consequently, this may lead to increased electron absorption at the metal surface. Overall, the number of electrons absorbed by the GEM3 metal surfaces is lower than the number of electrons transferred to the induction region. However, in the case of Config. 3 with an inner hole diameter greater than 20 $\mu$m, the electron loss at the GEM3 upper metal surface exceeds the number of electrons reaching the induction region. This behavior indicates a significant reduction in overall charge-collection efficiency at the induction plane.

The absolute number of trapped electrons increases with increasing inner hole diameter for all three configurations. The enhanced electron trapping resulting from an increased inner hole diameter is attributed to the expansion of the avalanche region within the GEM holes. Additionally, the increase in inner hole diameter may increase the lateral spread of electric field lines within the GEM holes, causing electrons to deviate towards the Kapton surface. 

Fig. \ref{fig:idce} shows the variation of collection efficiency (CE) as a function of inner hole diameter for the simulated GEM configurations. The CE initially increases with increasing inner hole diameter but tends to saturate with further enlargement. Moreover, it is observed that reduced pitch GEM detectors can achieve higher CE even at a low inner hole diameter. As the biconical profile becomes less pronounced, the distortion of the electric field line at the hole edges may decrease. This effect would allow more electrons to be guided into the holes, resulting in enhanced CE. However, beyond an optimum value, further increase in inner hole diameter may reduce field confinement, thereby negatively impacting CE.

Fig. \ref{fig:idr} shows the variation of $\sigma$ value as a function of inner hole diameter across all simulated GEM configurations. In Config. 1, the $\sigma$ values slightly increase with the increase in inner hole diameter. Config. 2 relatively maintains stable spatial spreading, while Config. 3 shows a slight improvement in electron sampling with increasing inner hole diameter. The minor variation of $\sigma$ values may result from the modified electric field lines due to the changes in the inner hole diameter.

\section{Impact of metal thickness \label{metalthickness}}

\begin{figure*}[htbp]
\centering
\includegraphics[width=.99\textwidth,height=6.5cm]{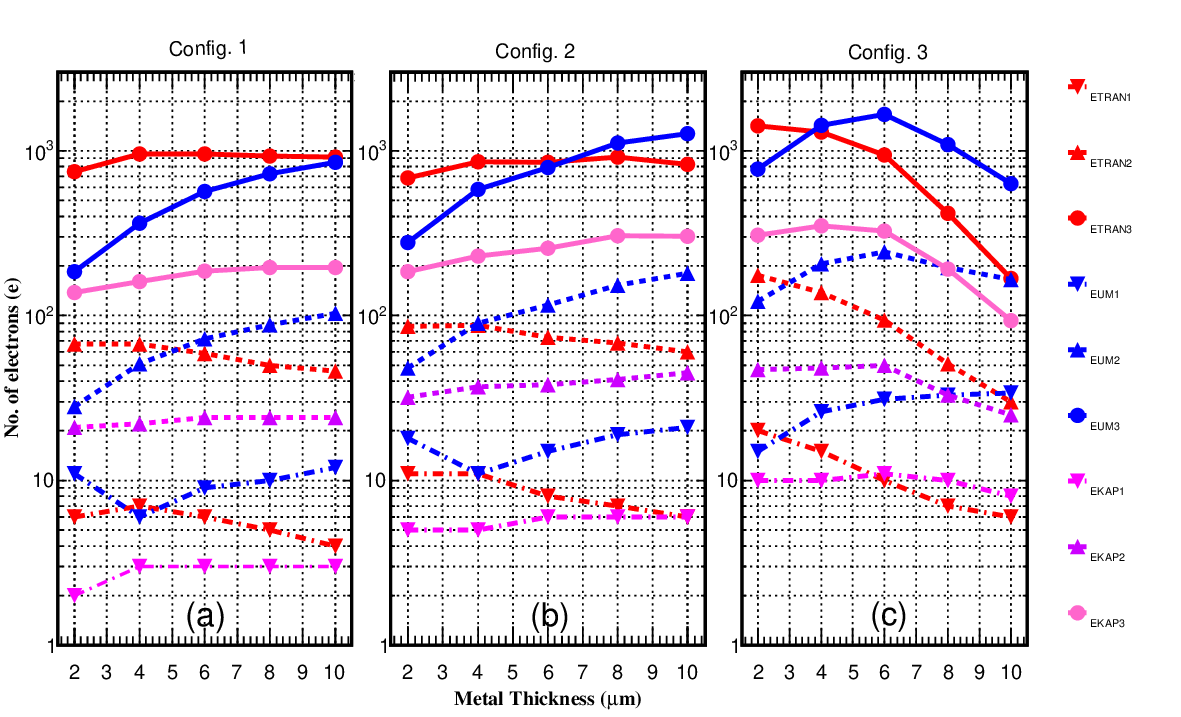}
\caption{Electron collections across Transfer, Upper Metal, and Kapton regions as a function of metal thickness for (a) Config. 1 at $\Delta V_{\mathrm{GEM}}$ = 360 V, (b) Config. 2 at $\Delta V_{\mathrm{GEM}}$ = 360 V, and (c) Config. 3 at$\Delta V_{\mathrm{GEM}}$ = 290 V. \label{fig15}}
\end{figure*}

Fig. \ref{fig15} shows the absolute number of electrons collected in the transfer regions, on the upper GEM metals, and inside the GEM holes as a function of metal thickness for all three distinct GEM configurations. As evident from the figure, Config. 1 achieves maximum electron collection in the Tran. 3 region at a metal thickness of 4 $\mu$m, beyond which the collection remains relatively stable. While electron collection in Tran. 1 and Tran. 2 regions of the same Config. 1 begins to decline after the 4 $\mu$m threshold. ETRAN1 in Config. 1 exhibits the same behavior as that of the standard single GEM as the metal thickness increases \cite{Mondal:2024ogs}. Config.~2 exhibits a similar trend of electron collection in the transfer regions as that of Config. 1. However, Config.~3 shows a consistent decrease in electron collection across all transfer regions as the metal thickness increases, indicating a reduction in charge collection efficiency at the induction electrode. The decrease in electron transmission across the corresponding transfer regions may result from an increase in metal area coverage. Additionally, the possible distortion of electric field lines caused by a larger metal area further contributes to this degradation. However, the observed consistency of electron collections in the Tran. 3 region of Configs. 1 and 2 may be attributed to the suppression of geometrical and field line effects due to a relatively wider transmission path.

Configs. 1 and 2 demonstrate a consistent increase in electron absorption at the upper metal surfaces with increasing metal thickness. However, an initial decrease in EUM1 is observed from 2 $\mu$m to 4 $\mu$m. In contrast, Config. 3 shows a non-monotonic trend of electron absorption across the upper GEM metals, where absorption initially increases to a maximum and then decreases with further increase in metal thickness. The initial increase observed in Config. 3 is possibly due to the larger metal thickness to hole depth ratio. However, with further increase in metal thickness, the possibility of electron absorption at the lower metal layers increases. This enhanced absorption reduces the seed electrons available for electron multiplication, thereby decreasing the avalanche rate in subsequent GEM holes. Consequently, fewer electrons may reach the upper metal layers, resulting in reduced absorption. This observation is supported by the fact that the reduction in upper metal absorption is not compensated by increased transmission or trapping.  Furthermore, in almost all three GEM configurations, the thicker metals exhibit relatively higher absorption than electron transmission.

In Configs. 1 and 2, it is observed that the electron lost due to trapping increases with increasing metal thickness. The deeper drift path within the GEM holes likely increases the possibility of trapping. In contrast, Config. 3 shows the reduction in these losses beyond a metal thickness of 6 $\mu$m. This behavior may also result from the overall decrease in avalanche rates across the system. 

\begin{figure*}[h!]
    \centering
    \begin{subfigure}[b]{0.48\textwidth} 
        \centering
        \includegraphics[width=\textwidth, height=5cm]{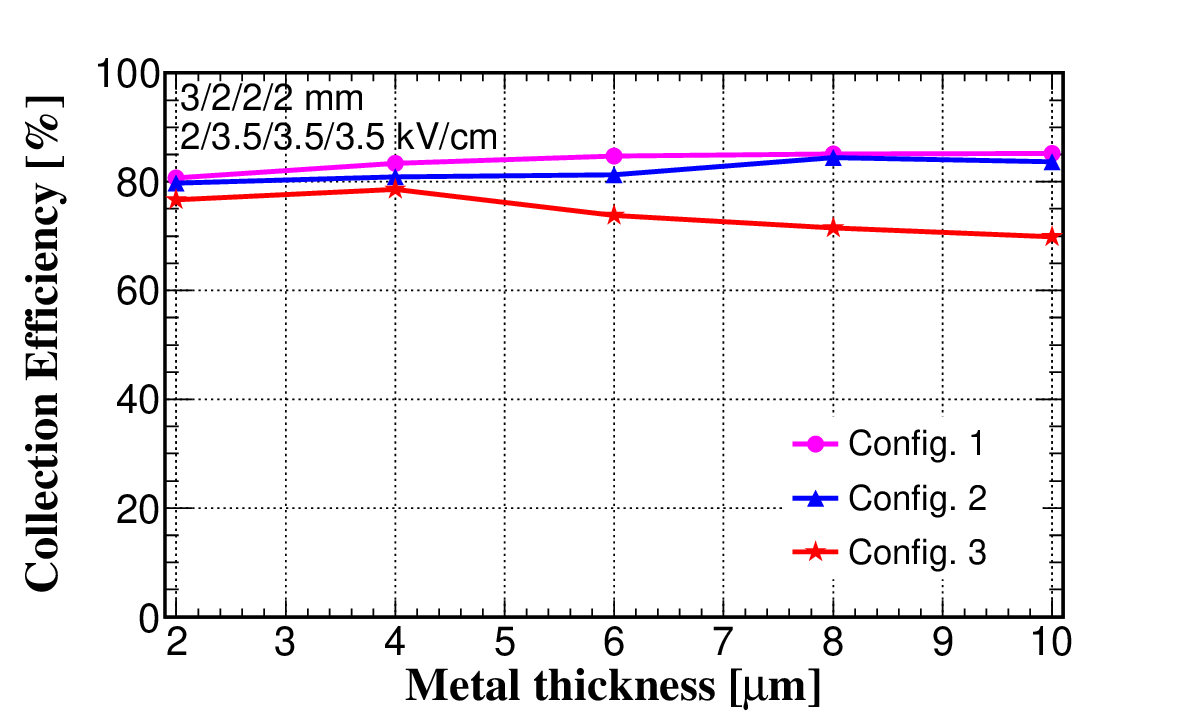}  
        \caption{}
        \label{fig:mtce}
    \end{subfigure}
    \hfill
    \begin{subfigure}[b]{0.48\textwidth}  
        \centering
        \includegraphics[width=\textwidth, height=5cm]{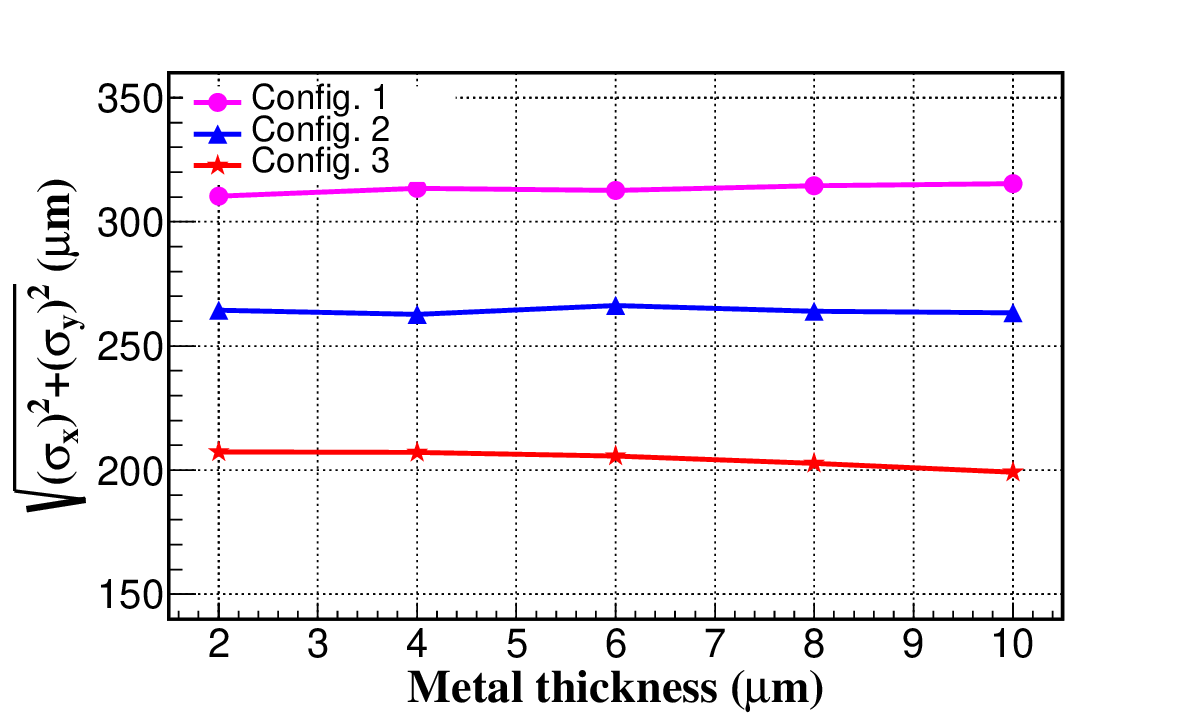}
        \caption{}
        \label{fig:mtr}
    \end{subfigure}

    \caption{(a) Collection efficiency as a function of metal thickness for three distinct GEM configurations. (b) Variation of  \(\sigma = \sqrt{\sigma_{x}^2 + \sigma_{y}^2}\) with metal thickness. The drift and transfer electric fields are set to 2 kV/cm and 3.5 kV/cm, respectively.}
    \label{fig:10}
\end{figure*}

Fig. \ref{fig:mtce} shows the variation of collection efficiency (CE) as a function of metal thickness. Configs. 1 and 2 exhibit nearly consistent CE with increasing metal thickness. However, Config. 3 shows degradation in CE beyond 4 $\mu$m. This degradation may arise due to its relatively higher ratio of metal width to outer hole diameter. Additionally, the increase in metallic thickness may also contribute to the observed degradation. 

Figure \ref{fig:mtr} shows that the $\sigma$ values are consistent, particularly in Configs. 1 and 2 as the metal thickness increases. It indicates a negligible impact of metal thickness on the spread of electrons. However, a slight improvement in electron sampling is observed for Config. 3 as the metal thickness increases. This slight improvement may result from collimated electrons guided by thicker metals.

\section{Impact of Kapton thickness \label{kaptonthickness}}

\begin{figure*}[t!]
\centering
\includegraphics[width=.99\textwidth,height=6.5cm]{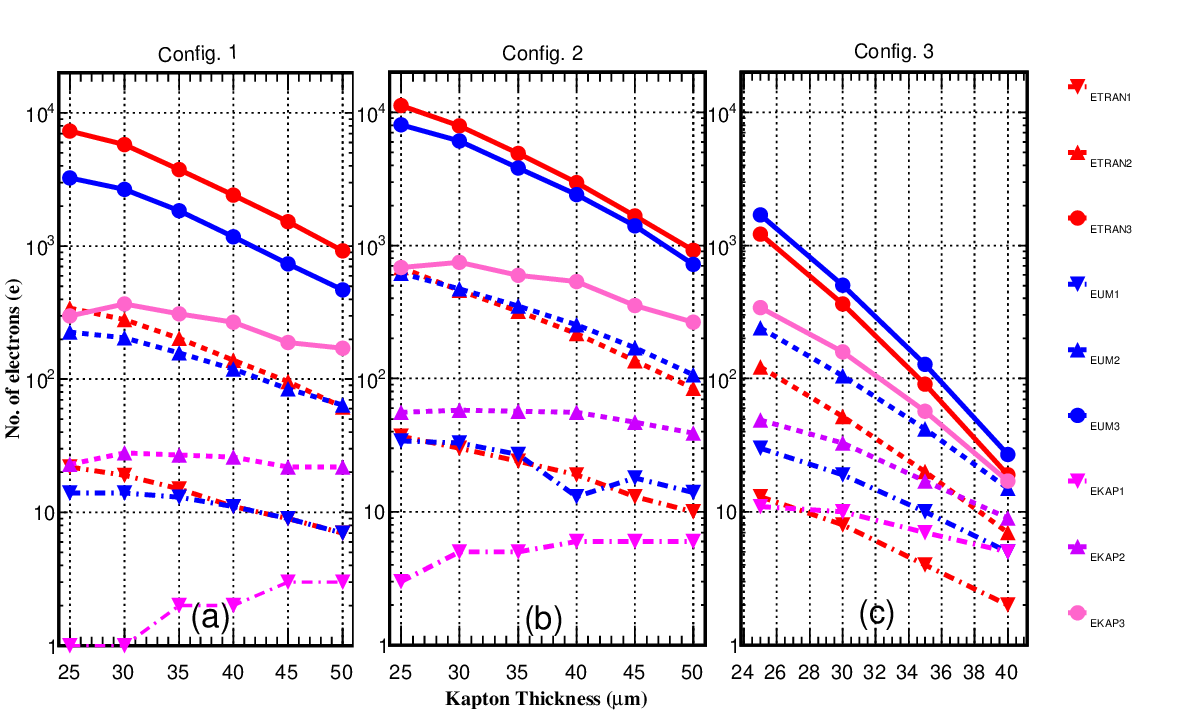}
\caption{Electron collections across Transfer, Upper Metal, and Kapton regions as a function of Kapton thickness for (a) Config. 1 at $\Delta V_{\mathrm{GEM}}$ = 360 V, (b) Config. 2 at $\Delta V_{\mathrm{GEM}}$ = 360 V, and (c) Config. 3 at $\Delta V_{\mathrm{GEM}}$ = 290 V. \label{fig17}}
\end{figure*}

Figure~\ref{fig17} shows the absolute number of collected electrons in the transfer regions, on the GEM electrodes, and within the GEM holes as a function of Kapton thickness. It is observed that number of transferred electrons decreases with an increase in Kapton thickness. Among them, Config.~2 shows a relatively higher electron transfer through the transfer regions than Configs.~1 and~3. The observed reduction in electron transmission with increasing Kapton thickness may be attributed to the corresponding decrease in electric field strength, given that the field strength is inversely proportional to the foil thickness at constant GEM potential. Moreover, the results indicate that GEMs with reduced pitch should employ thinner Kapton foils than the standard thickness (50 $\mu$m) to achieve efficient electron transport.

\begin{figure*}[t!]
    \centering
    \begin{subfigure}[b]{0.48\textwidth} 
        \centering
        \includegraphics[width=\textwidth, height=5cm]{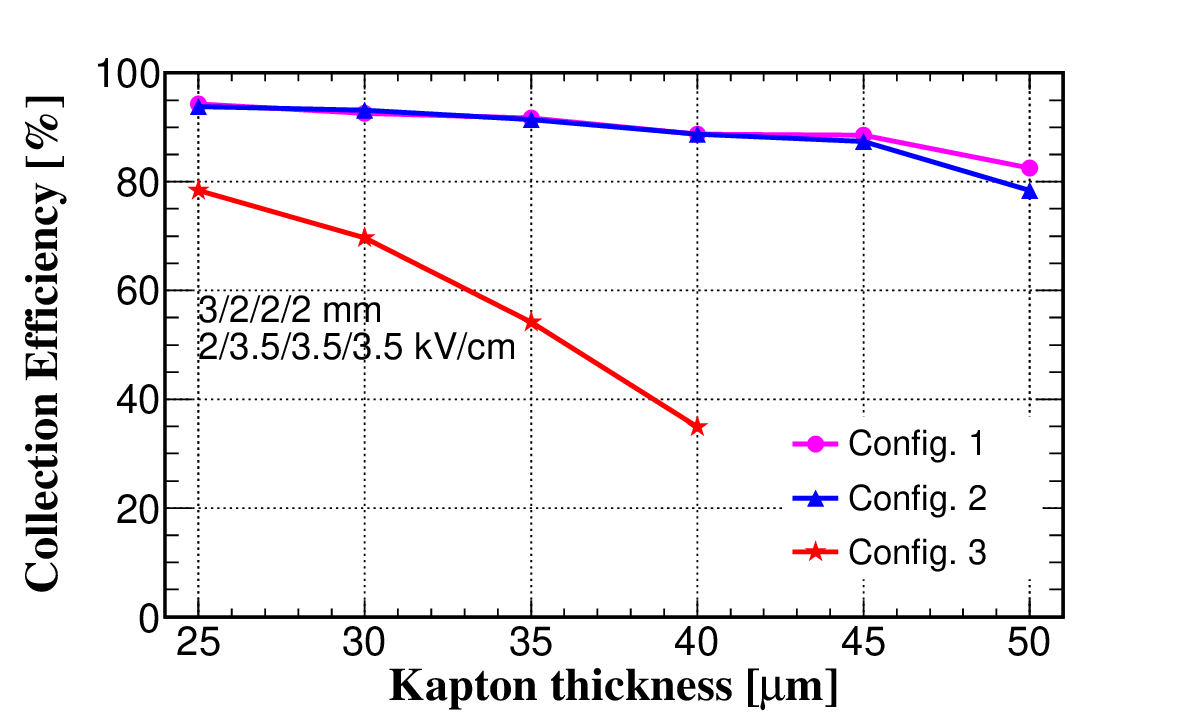}  
        \caption{}
        \label{fig:cekt}
    \end{subfigure}
    \hfill
    \begin{subfigure}[b]{0.48\textwidth}  
        \centering
        \includegraphics[width=\textwidth, height=5cm]{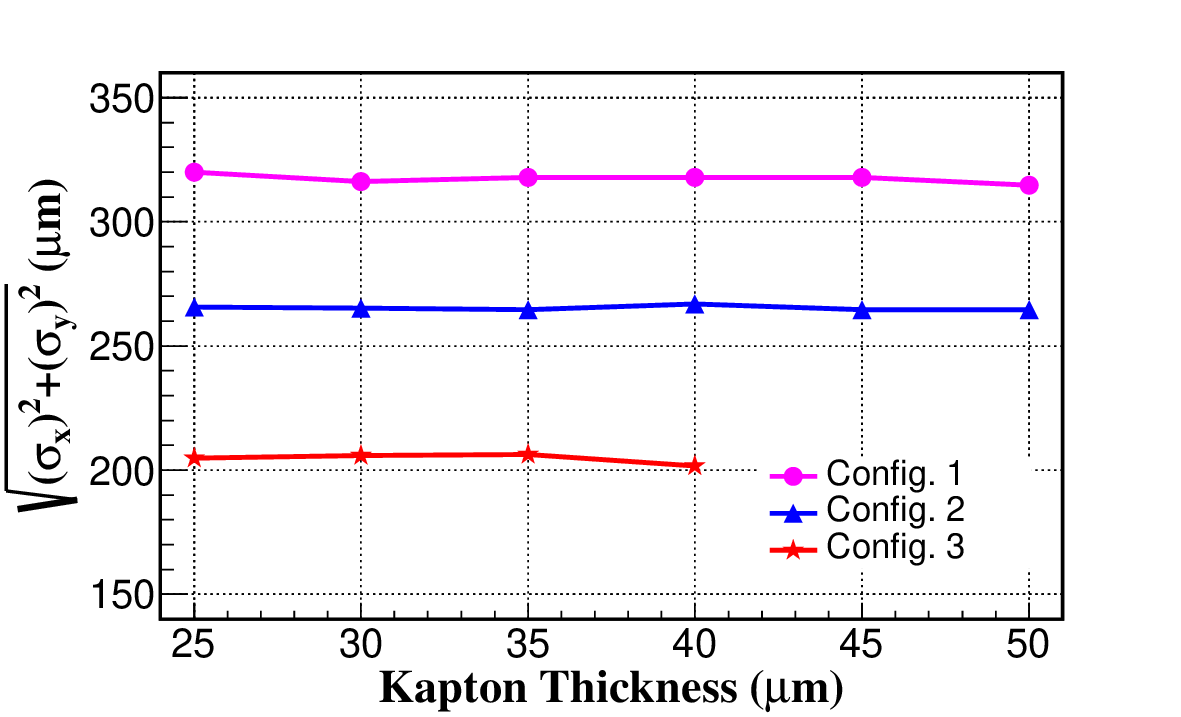}
        \caption{}
        \label{fig:ftr}
    \end{subfigure}

    \caption{(a) Collection efficiency as a function of Kapton thickness for three distinct GEM configurations. (b) Variation of  \(\sigma = \sqrt{\sigma_{x}^2 + \sigma_{y}^2}\) with Kapton thickness. The drift and transfer electric fields are set to 2 kV/cm and 3.5 kV/cm, respectively.}
    \label{fig:10}
\end{figure*}

Electron absorption by the upper metal surfaces decreases with increasing Kapton thickness across all three configurations. The metallic absorption is estimated to be more significant in Config. 2. In Config. 3, the loss of electrons due to metal absorptions is more significant than the transferred electrons. The increase in electron absorption with reduced Kapton thickness is likely due to the curved electric field lines near the metal layers, which tend to deflect outgoing electrons toward the metal surfaces. Moreover, reducing Kapton thickness shortens the avalanche path length but simultaneously increases the avalanche rates, resulting in greater absorption at the metal surfaces.

Electron trapping decreases with increasing Kapton thickness, particularly within the holes of GEM2 and GEM3 across all three configurations. In contrast, an increase in Kapton thickness leads to enhanced electron trapping in the GEM1 holes of Configs. 1 and 2. The observed decrease in electron trapping with increasing Kapton thickness is likely due to low avalanche effects. An exception is observed in Config. 3, where the number of electrons transferred to the Tran. 1 region becomes lower than the electrons trapped in GEM1 with increasing Kapton thickness. This behavior indicates that thicker ($>$ 25 $\mu$m), small-pitch GEM (60 $\mu$m) exhibits enhanced electron trapping within the GEM1 holes. 

Fig.~\ref{fig:cekt} shows the variation of collection efficiency (CE) as a function of Kapton thickness for the simulated GEM configurations. Increasing Kapton thickness results in degradation of CE. Config. 3 shows comparatively higher degradation. The decrease in collection efficiency may arise from the increased hole aspect ratio (Kapton thickness to hole diameter) and reduced electrofocusing. Since Config. 3 features, including dense hole patterns, weakening of electrofocusing, and the effect of the higher aspect ratio, become more dominant, leading to severe degradation.

Fig.~\ref{fig:ftr} show that the $\sigma$ remains constant as the Kapton thickness increases. The stability of $\sigma$ across different Kapton thicknesses suggests that transverse diffusion in the Tran.~3 region is largely independent of the Kapton thickness. It also indicates that the primary contribution to the improved electron sampling resulting from variations in physical parameters originates from the reduction in pitch dimension.

\begin{figure*}[t!]
\centering
\includegraphics[width=.99\textwidth,height=6.5cm]{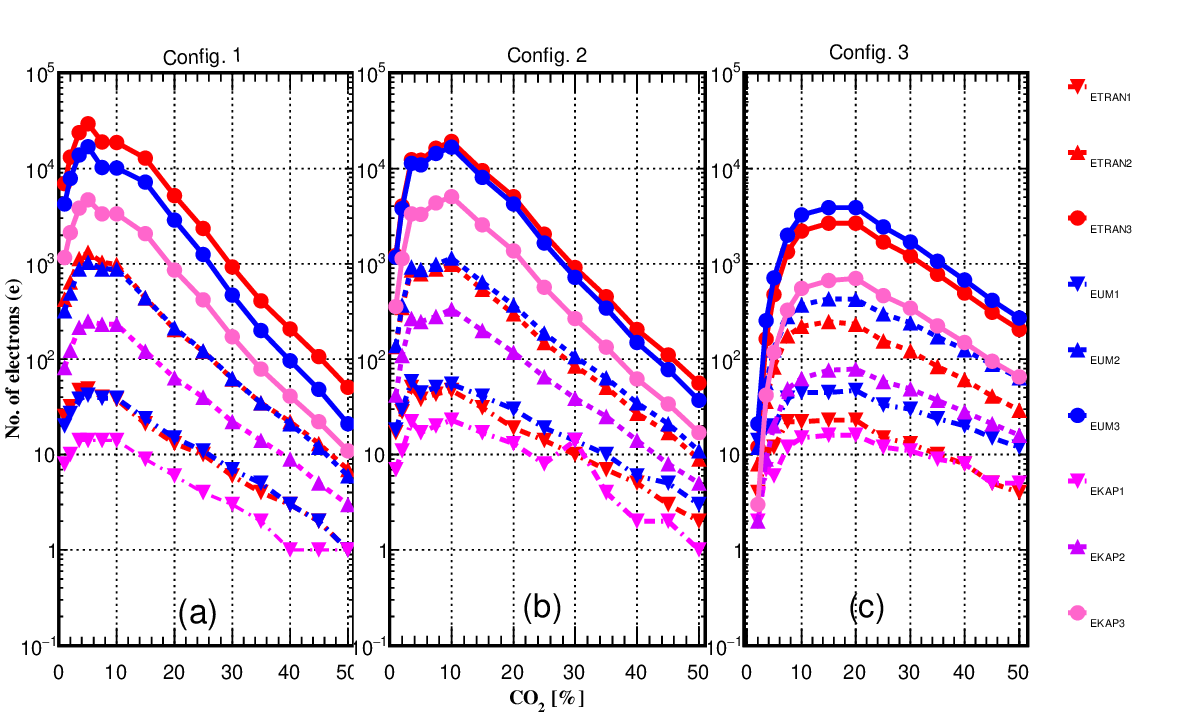}
\caption{Electron collections across Transfer, Upper Metal, and Kapton regions as a function of CO$_{2}$ concentrations [\%] for (a) Config. 1 at $\Delta V_{\mathrm{GEM}}$ = 360 V, (b) Config. 2 at $\Delta V_{\mathrm{GEM}}$ = 360 V, and (c) Config. 3 at$\Delta V_{\mathrm{GEM}}$ = 290 V. \label{fig19}}
\end{figure*}

\begin{figure*}[t!]
\centering
\includegraphics[width=.5\textwidth]{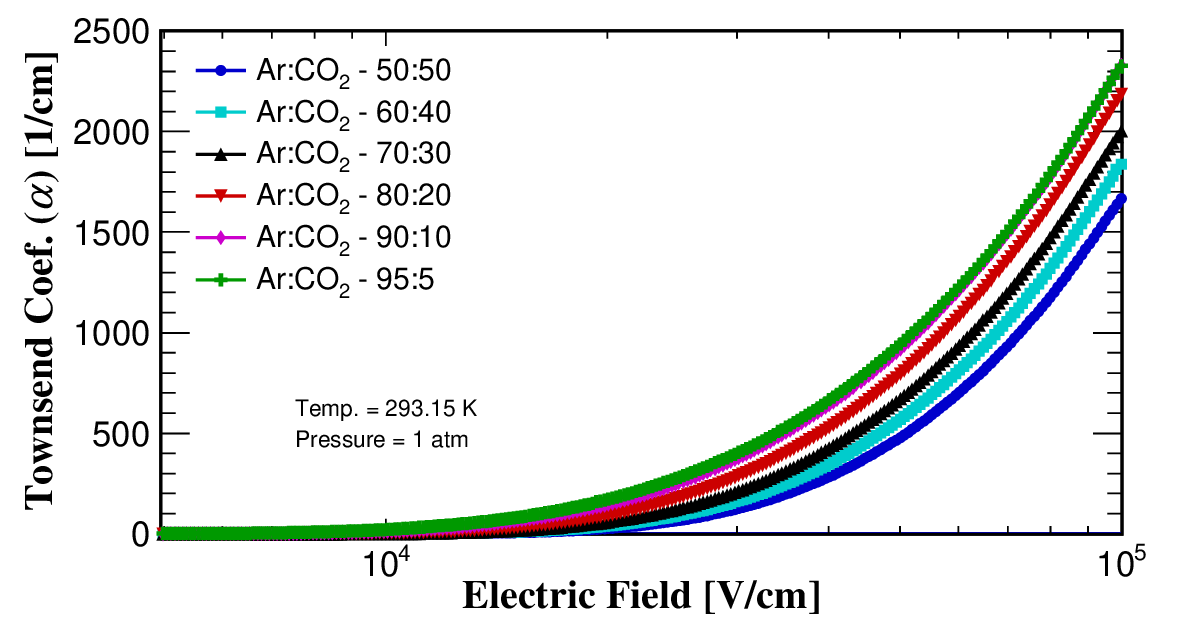}
\caption{Magboltz simulation: Townsend coefficient incorporating Penning effect vs. electric field at 1 atm and 293.15 K  \label{fig:townsend}}
\end{figure*}

\section{Impact of CO$_{2}$ concentration \label{gasconcentration}}

Fig.~\ref{fig19} shows the absolute number of collected electrons across the transfer regions, the upper GEM electrodes, and within the GEM holes as a function of CO$_{2}$ concentration [\%] for all three GEM configurations. As the concentration of CO$_{2}$ increases, the concentration of Argon is proportionally reduced to maintain the gas mixture ratio. The Penning transfer ratio (r$_{p}$) is adjusted accordingly based on the CO${_2}$  content to account for the Penning effect, and its specific values are sourced from Refs. \cite{Gupta:2025yrv, Sahin:2014haa}. The figure shows that as the CO$_{2}$ concentration increases, the number of electrons collected in the transfer regions across the simulated GEM configurations initially increases, reaches a maximum, and then declines. A similar trend has also been observed for single GEMs \cite{Gupta:2025yrv}. Additionally, the maximum collection point shifts toward higher CO$_{2}$ concentrations as the configuration's pitch size decreases. For instance, the maximum electron transfer occurs around 5\% CO$_{2}$ in Config. 1, around 10\% in Config. 2, and approximately 20\% in Config. 3. The gas gain (G) over a path length from r$_{1}$ to r$_{2}$ under the influence of an electric field ($E$) can be expressed as \cite{Davydov:2004bv}:

\begin{equation}
G = \exp\left( \int_{r_1}^{r_2} \alpha(r) \, dr \right),
\end{equation}

where $\alpha(r)$ is the first Townsend coefficient, defined as the number of ionizing collisions per cm. $\alpha(r)$ is the function of reduced electric field strength $(E/p)$, where p is pressure. As the CO$_{2}$ concentration decreases, $\alpha$ correspondingly increases at the given electric field, as supported by Fig.~\ref{fig:townsend}. Such behavior has also been observed experimentally \cite{Sharma:1993wp}. The increase in $\alpha$ leads to enhanced gas gain, resulting in a greater number of electrons being collected in the transfer regions. Moreover, the increase in Argon content enhances secondary ionization, thereby further increasing electron collection. However, at lower CO$_{2}$ concentration, the observed reduction in electron transmission may be attributed to the lower collection efficiency. Additionally, at low CO$_{2}$ concentration, the penning transfer ratio (r$_{p}$) is significantly low, resulting in reduced secondary ionization leading to lower electron collection in the transfer regions. The reduced electron collection observed in Config. 3, compared to Configs. 1 and 2 may be attributed to its operation at a lower GEM voltage ($\Delta V_{\mathrm{GEM}}$= 290 V). While the enhanced electron transfer at high CO$_{2}$ concentration in Config. 3 may arise from reduced electron diffusion and improved transparency of the GEM foils.

\begin{figure*}[h!]
    \centering
    \begin{subfigure}[b]{0.49\textwidth} 
        \centering
        \includegraphics[width=\textwidth, height=5cm]{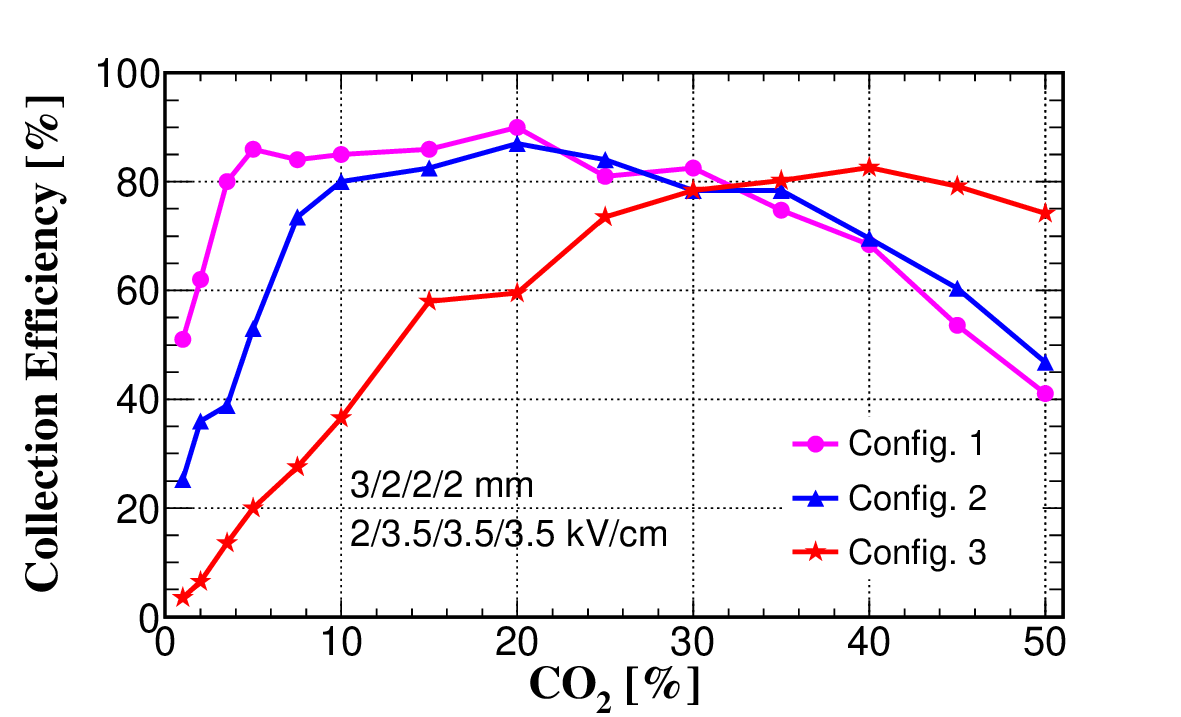}  
        \caption{}
        
        \label{fig17a}
    \end{subfigure}
    \hfill
    \begin{subfigure}[b]{0.49\textwidth}  
        \centering
        \includegraphics[width=\textwidth, height=5cm]{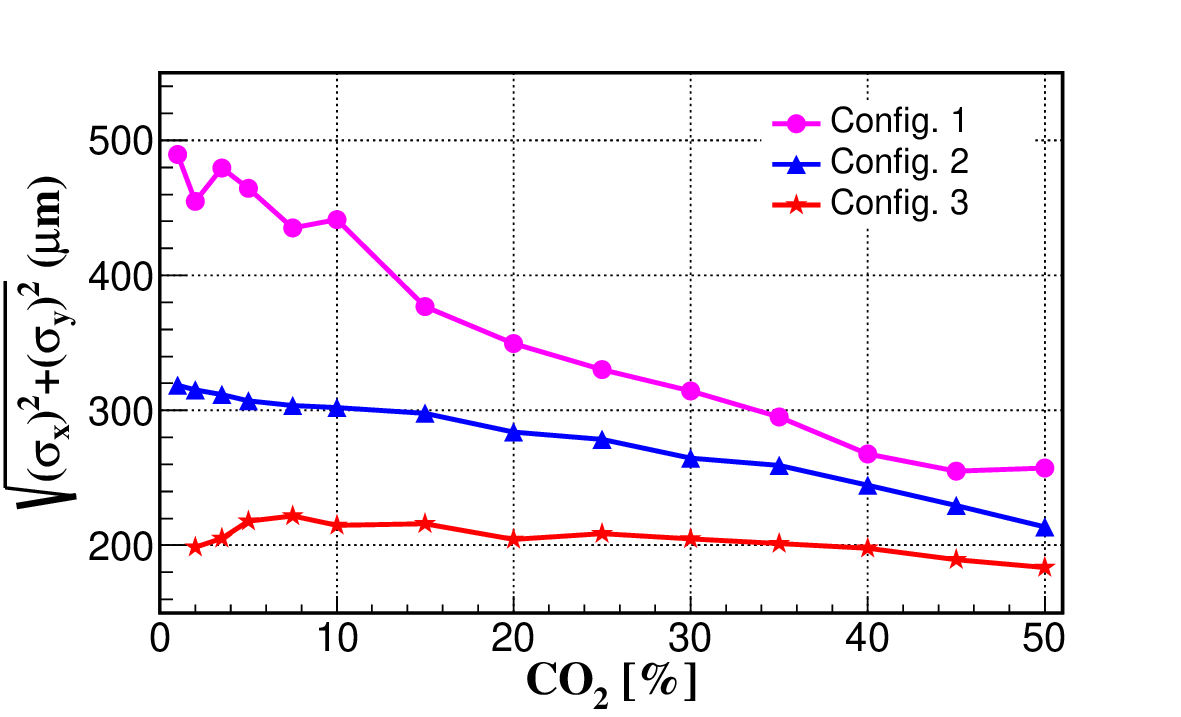}
        \caption{}
        \label{fig17b}

    \end{subfigure}

    \caption{(a) Collection efficiency as a function of CO$_{2}$ [\%] for three distinct GEM configurations. (b) Variation of  \(\sigma = \sqrt{\sigma_{x}^2 + \sigma_{y}^2}\) with CO$_{2}$ [\%]. The drift and transfer electric fields are set to 2 kV/cm and 3.5 kV/cm, respectively.}
    \label{fig20}
    
\end{figure*}

\begin{figure*}[htbp]
\centering
\includegraphics[width=.5\textwidth]{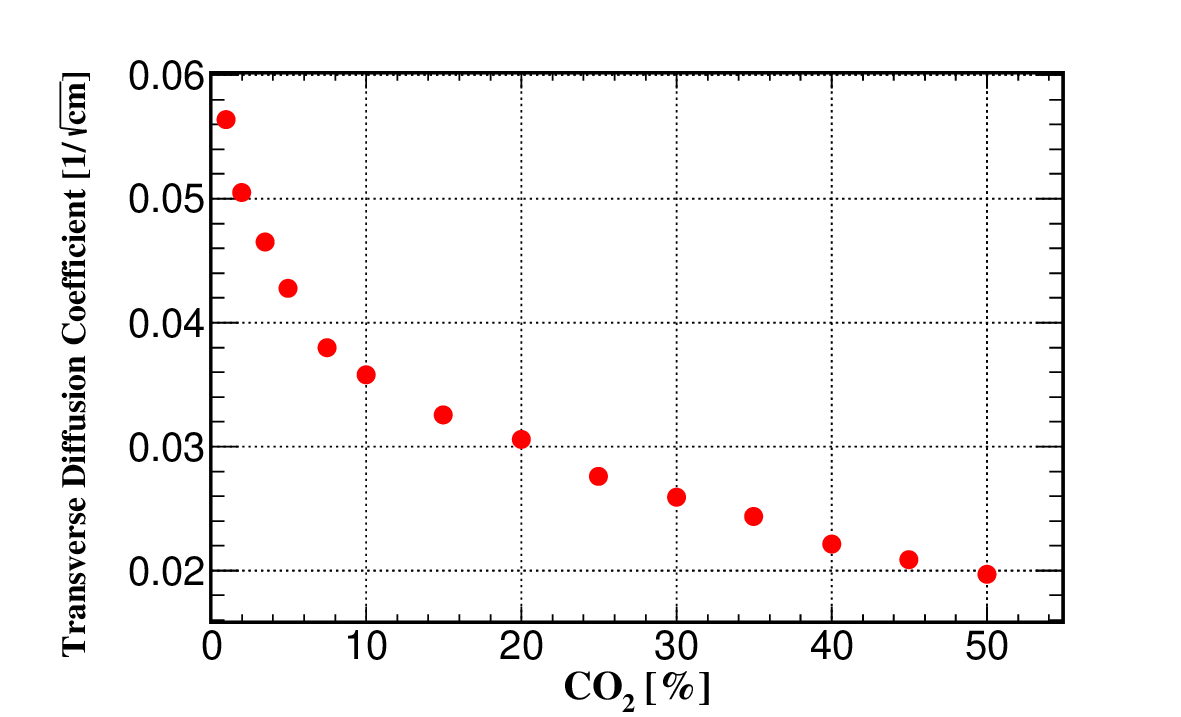}
\caption{Variation of the transverse diffusion coefficient of electrons as a function of CO$_{2}$ concentration [\%] at an electric field of 3.5 kV/cm. \label{magboltz}}
\end{figure*}

It is observed that the metallic absorption follows the same non-monotonic trend as transferred electrons. Config. 2 exhibits relatively higher absorption compared to Configs. 1 and 3. Furthermore, Config. 3 shows that the electron transfer is less efficient, as a larger number of electrons are lost due to metallic absorption. At low CO$_{2}$ concentrations, avalanche electrons exhibit reduced drift velocity and increased diffusion \cite{Assran:2011ug}, causing a greater fraction of electrons to be collected at the metal surfaces. Variation in the electric field distribution could be a possible reason for the differences in metallic absorption observed across the simulated GEM configurations.

The electron lost due to trapping follows an established non-monotonic trend for all GEM configurations. At low CO$_{2}$ concentrations, electrons can drift freely with sufficient average energies, making them less susceptible to being trapped within the GEM holes. However, as CO$_{2}$ concentration increases, the enhanced collision frequency causes greater energy loss, thereby increasing the probability of being trapped in the low-field region of holes. The observed consistent trend across the simulated GEM configurations indicates a common influence of physical dimensions, electric field distribution, and CO$_{2}$ concentrations on electron trapping.

Fig.~\ref{fig17a} shows that the collection efficiency (CE) initially increases with CO$_{2}$ concentration, remains stable over a certain range, and then decreases with further addition of CO$_{2}$ concentration. It is also evident that Configs. 2 and 3 exhibit CE greater than over 80 $\%$ at a narrower CO$_{2}$ concentration range. However, these configurations can achieve higher CE than Config. 1 at higher CO$_{2}$ concentrations. These CE behaviors are also reflected in the number of collected electrons, as shown in Fig.~\ref{fig19}. At low CO$_{2}$ concentrations, transverse diffusion is high, which may divert the incoming electrons towards the metal surfaces. Increasing the CO$_{2}$ concentration enhances inelastic collisions, reducing mean electron energy and thereby suppressing diffusion. This phenomenon can create an optimal range in which the focusing field lines can effectively guide electrons into the GEM holes, thereby enhancing the CE. However, at higher CO$_{2}$ concentrations, increased electron attachment and reduced drift velocity lead to significant electron loss in the drift region and at the metallic surfaces, ultimately degrading CE.

Figure \ref{fig17b} shows the variation of $\sigma$ as a function of CO$_{2}$ concentration for the investigated GEM configurations. As CO$_{2}$ concentration increases, the $\sigma$ values in Configs. 1 and 2 decrease. In contrast, Config. 3 shows an increase in $\sigma$ up to a CO$_{2}$ concentration of 7.5\%, followed by a reduction. The declining rate in $\sigma$ values is least significant in Configs. 2 and 3 compared to Config. 1. The decreased $\sigma$ exhibited by reduced pitch GEM detectors across a wide range of CO$_{2}$ concentration compared to standard Config. 1 highlights improvement in electron sampling due to pitch size reduction. The reduction in $\sigma$ across the simulated GEM configuration with increasing CO$_{2}$ concentration can be explained by the diffusion properties of the gas mixture. 

The resultant standard deviation (\(\sigma = \sqrt{\sigma_{x}^2 + \sigma_{y}^2}\)) follows a similar behavior to the transverse diffusion coefficient (D$_{T}$) given by Eq. \ref{eqndiffusion}. Magboltz computes transverse spread as $\sigma_{T} = D_{T}\sqrt{s}$ \cite{Sauli:1977mt, GarfieldPP_UserGuide_2025}, where $D_{T}$ is evaluated in unit of $\mathrm{cm}^{-1/2}$. The coefficient D$_{T}$ is obtained using displacement formalism \cite{Biagi:1999nwa}, which incorporates the effects of electric field, pressure, magnetic field, and gas composition. Fig. \ref{magboltz} shows the evolution of D$_{T}$ with increasing CO$_{2}$ concentration at a fixed electric field of 3.5 kV/cm. As evident from the figure, D$_{T}$ decreases with increasing CO$_{2}$ concentration, indicating corresponding reduction in $\sigma$ and $\sigma_T$. Furthermore, it is observed that the reduction in D$_{T}$ has relatively lesser impact on $\sigma$ for the reduced pitch GEM detector as evident from Fig. \ref{fig17b}. These results highlight that Configs. 2 and 3 benefit more substantially from enhanced field focusing, enhancing the spatial confinement of the electron cloud.

\section{Summary and Conclusions \label{summary}}

This article presents a detailed analysis and comparison of electron loss, transmission across different regions, collection efficiency, and spatial spreads for three distinct GEM configurations: Config.1 (140 $\mu$m pitch), Config. 2 (90 $\mu$m pitch), and Config. 3 (60 $\mu$m pitch). Motivated by experimental and simulation findings that reduced-pitch-sized (90 $\mu$m and 60 $\mu$m) GEM configurations provide enhanced gain, collection efficiency, and spatial resolution, we conducted a comprehensive analysis of these triple GEM detectors. At first, the simulation framework was validated against existing simulation and experimental results. Following validation, the impact of GEM voltage ($\Delta V_{\mathrm{GEM}}$), outer hole diameter, inner hole diameter, metal thickness, Kapton thickness, and CO$_{2}$ concentration was investigated.

The simulation reveals that Config. 2 exhibits electron losses and transmission similar to those of Config. 1 when $\Delta V_{\mathrm{GEM}}$ is increased. Config. 2 achieves approximately 12\% lower electron spread compared to Config. 1. In contrast, Config. 3 shows enhanced avalanche rates and a subsequent increase in electron collections at different regions of the simulated GEM structure. Notably, at $\Delta V_{\mathrm{GEM}}$ of 320 V, Config. 3 transmits nearly about 10$^{2}$ times higher electrons than Configs. 1 and 2 in the induction region. Moreover, Config. 3 reduces the electron spread by nearly 30\% relative to Config. 1 and 26\% relative to Config. 2.

Furthermore, the simulation results indicate that electron losses and transmission are significantly affected by variations in the investigated geometrical parameters and gas compositions. For instance, within the simulated parameter space, increasing outer hole diameter shows non-monotonic (rise and fall) behavior in electron losses and transmission for Configs. 1 and 3, while Config. 3 shows a predominantly decreasing trend. Increasing the inner hole diameter enhances electron collection, whereas increasing the Kapton thickness reduces it. For Configs. 1 and 2, electron transmission remains stable beyond a metal thickness of 4 $\mu$m, whereas Config. 3 shows a decreasing trend across the same range. The electron losses with increasing metal thickness also follows the non-monotonic behavior. Additionally, increasing the CO$_2$ concentration results in a rise and fall behavior of electron collection across all configurations. 

The investigation of collection efficiency reveals that the maximum CE can be achieved in Config. 1, when $\Delta V_{\mathrm{GEM}}$ is greater than 360 V, the outer hole diameter is around 55 - 65 $\mu$m, the inner hole diameter is around 50 - 55 $\mu$m, the metal thickness is around 2 - 10 $\mu$m, the Kapton thickness is less than 50 $\mu$m, and the CO$_2$ composition is between 5 -30$\%$. For Config. 2, maximum CE is obtained when $\Delta V_{\mathrm{GEM}}$ is greater than 370 V, the outer hole diameter is around 40 - 55 $\mu$m, the inner hole diameter is around 40 - 50 $\mu$m, the metal thickness is around 2 - 10 $\mu$m, the Kapton thickness is less than 50 $\mu$m, and the CO$_2$ composition is between 10 -35$\%$. For Config. 3, maximum CE is obtained when $\Delta V_{\mathrm{GEM}}$ is greater than 300 V, the outer hole diameter is around 25 - 30 $\mu$m, the inner hole diameter is around 22.5 - 27.5 $\mu$m, the metal thickness is less than 4 $\mu$m, the Kapton thickness is around 25 $\mu$m, and the CO$_2$ composition is between 30 -45$\%$.

Furthermore, the simulation results indicate that electron spread is primarily influenced by the gas composition, while remaining relatively consistent across variations in the other investigated parameters. It is worth noting that in all the simulation cases, Config. 3 consistently delivers better electron sampling, followed by Config. 2, and then Config. 1. This trend indicates that the dimension of pitch size plays a crucial role in the spread of electron cloud at the induction electrode.

The favorable results, particularly the enhanced electron transfer, adequate collection efficiency, and improved electron sampling resulting from the reduced pitch dimension, suggest that these detectors can be a suitable choice for future high-energy physics experiments. Furthermore, the comprehensive analysis of variations in physical dimensions and gas composition provides valuable insights for fabrication and determination of suitable operating conditions. However, the intrinsic limitations of Garfield++ and the exclusion of readout electronics could cause discrepancies between simulated and expected experimental results. Moreover, to fully conclude the improvement in GEM performance due to pitch size reduction, a detailed analysis of ion backflow is essential, which we plan to explore in the future. Future experimental investigations based on the simulated parameters will help validate and reinforce the improved performance associated with reduced hole pitch size.

\hfill

\section*{Acknowledgements}
AK sincerely acknowledges the financial support from the Institute of Eminence (IoE), BHU grant - 6031. RG and SS acknowledge the financial support from UGC under the research fellowship scheme in central universities.

\section*{Data availability}
Data will be made available on request.

\end{document}